\documentclass[twocolumn,nofootinbib,floatfix,superscriptaddress]{revtex4}        
\usepackage{graphicx}
\usepackage{epsfig}
\usepackage{bm}
\usepackage[T1]{fontenc}
\usepackage[latin9]{inputenc}
\usepackage{amssymb}
\usepackage{float}
\usepackage{amsmath}
\usepackage{dcolumn}
\usepackage{cancel}
\usepackage[colorlinks]{hyperref}
\usepackage[usenames,dvipsnames]{color}
\hypersetup{
    breaklinks=true,
    pdfstartview={FitH},    
    colorlinks=true,       
    linkcolor=blue,          
    citecolor=red,        
    filecolor=magenta,      
    urlcolor=blue,           
    anchorcolor=green,      
    linktocpage=true
}

\newcommand{\be}{\begin{equation}}
\newcommand{\ee}{\end{equation}}

\begin{document}

\title{Nonextensive entropies impact onto thermodynamics and phase structure of Kerr-Newman black holes}

\author{S. Ghaffari}
\email{sh.ghaffari@maragheh.ac.ir}
\affiliation{Research Institute for Astronomy and Astrophysics of Maragha (RIAAM),
University of Maragheh, P.O. Box 55136-553, Maragheh, Iran}

\author{G.~G.~Luciano}
\email{giuseppegaetano.luciano@udl.cat}
\affiliation{Department of Chemistry, Physics, Environmental and Soil Sciences, Escola Polit\`ecnica Superior, Universitat de Lleida, Av. Jaume
II, 69, 25001 Lleida, Spain}

\author{A. Sheykhi}
\email{asheykhi@shirazu.ac.ir}
\affiliation{Department of Physics,
College of Sciences, Shiraz University, Shiraz 71454, Iran}
\affiliation{Biruni Observatory, College of Sciences, Shiraz
University, Shiraz 71454, Iran}

\date{\today}
\begin{abstract}
Taking the nonextensive Tsallis and R\'enyi entropies into
account, we explore thermodynamic properties and phase transitions
of the Kerr-Newman black holes (KNBH) in the microcanonical and
canonical ensembles. We also compare our results with those
obtained by attributing the Bekenstein-Hawking entropy bound to
the mentioned black holes. Our analysis indicates that, similarly
to the standard Boltzmann picture, isolated KNBH in the
microcanonical approach are stable against axisymmetric
perturbations in both Tsallis and Rényi models. On the other hand,
in considering the case when the black holes are enveloped by a
bath of thermal radiation in the canonical treatment, the KNBH
based on the Tsallis and R\'enyi entropies can be stable for some
values of the entropy parameters, in contrast to the traditional
Boltzmann framework. For the case of R\'enyi entropy, we find that
a Hawking-Page transition and a first order small black hole/large
black hole transition can occur in a similar fashion as in
rotating black holes in an anti-de Sitter space. Finally, we
employ the Ruppeneir geometrothermodynamic technique to provide a
new perspective on studying the nature of interactions between
black hole microstructures, revealing a non-trivial impact of
nonextensive entropies.
\end{abstract}

\maketitle

\section{Introduction}
The study of black holes and their thermodynamic properties have
been a subject of great interest and significance in theoretical
physics since the pioneering works of Bekenstein and Hawking on
black hole radiation in  1970's.
The standardized and simplified method for describing the
thermodynamics of black holes is based on Bekenstein-Hawking
entropy, derived from classical statistical mechanics and quantum
field theory, which has provided significant insight into the
thermodynamic properties of black
holes.~\cite{Bekenstein,Bek2,Hawking;1974,Hawking1,Hawking2,Bardeen,Gibbs}.
Despite the numerous achievements in the field of thermodynamics
of black holes, some important issues remain unsolved, such as the
nonextensive nature of black holes and the corresponding issue of
the thermodynamic stability. The thermodynamic and stability
problem of the different classes of black holes have been largely
studied in various
literature~\cite{Landsberg1,Landsberg2,Tsallis,Pavon,Maddox,Pesci,Davies,Kaburaki1,Kaburaki1bis,Kaburaki2,Czinner,Czinner2,Czinner3,Czinner4,Azreg}.

In standard statistical descriptions, interactions over long distances are often
considered insignificant. However, in the presence of strong gravitational fields,
such as black holes, these long-range interactions cannot be ignored.
As a result, the Boltzmann-Gibbs statistics may not be the most appropriate approach
for defining entropy in these strongly gravitating systems. This shortcoming was first noted by Gibbs, who pointed out that systems
with divergent partition functions, including those affected by gravitation, fall outside the scope of Boltzmann-Gibbs theory~\cite{Gibbs}.

In recent decades, the theory of thermodynamics based on
nonextensive and/or nonadditive entropy notion
has made significant progress. In the seminal work~\cite{Tsallis},
it was realized that the thermodynamic entropy of
black holes should be appropriately generalized
to recover thermodynamic extensivity. The ensuing modified Tsallis-Cirto entropy is in the form~\cite{Tsallis}
\begin{equation}
\label{ST1}
S_T=\gamma\left(\frac{ A}{A_0}\right)^\delta\,,
\end{equation}
where $A$ is horizon area of black hole, while $A_0$
the Planck area. The positive entropic parameter $\delta$ and the normalization factor $\gamma$ quantify departure from Bekenstein-Hawking entropy, with the limit case being given by $\delta=\gamma=1$.\footnote{It is worth noting that Eq.~\eqref{ST1} does not provide the most general definition of Tsallis entropy, but is rather the special member of the two-parameter class $S_{q,\delta}=\left(\log_q W\right)^\delta$ for $q=1$, where $W$ is the number of microstates of the system~\cite{Tsallis}.}

Along the same conceptual lines, several other parametric extensions of the Boltzmann-Gibbs statistical entropy formula have been introduced, which are more suitable for describing complex systems with long-range interactions. Among these, the R\'enyi entropy is constructed by taking the formal logarithm of the Bekenstien-Hawking entropy $S_B$ as follows
\begin{equation}
\label{SR1}
S_R=\frac{1}{\lambda}\ln\left(1+\lambda S_{B}\right)\,,
\end{equation}
where the deformation parameter
$\lambda$  is connected to the nonextensive and nonlocal nature of $S_R$, in such a way that $S_R\rightarrow S_B$ in the limit $\lambda\rightarrow0$.

The generalized entropies~\eqref{ST1} and~\eqref{SR1}  are at the core of extended statistics, which have significant implications for cosmological
settings~\cite{Davies,Kaburaki1,Kaburaki1bis,Kaburaki2,Czinner,Czinner2,Czinner3,Czinner4,Azreg,Biro,Nojiri-non1,Nojiri-non2,Luciano1,Nunes,Komatsu,Nojiri-non3,Ghaff1,Tavayef:2018xwx,Saridakis:2018unr,Nojiri:2019skr,Jizba,Vagnozzi,Morad2,LucianoGen,Boulka,Morad3} and
modified gravity~\cite{Capoz1,Capoz2,Capoz3,Capoz4,Capoz5,Capoz6,LucianoBar}.
Recently, new effort has been devoted to understand the impact of nonextensivity on the phase structure of black holes~\cite{Luciano,Luciano:2023bai,Jawad,Yassine,Cimidiker:2023kle,Morad1}. Specifically, in~\cite{Luciano} it has been found that
Tsallis prescription affects the small-large black holes phase-transitions of charged and static anti-de Sitter (AdS) black holes, while combined effects of nonextensivity and generalized uncertainty relations have been explored in~\cite{Cimidiker:2023kle} for the simplest spherically symmetric Schwarzschild black holes.
Despite such non-trivial results, further research is needed to fully grasp the implications of nonextensive entropies in the study of black holes. In particular, to account for astrophysical reality, the case of charged and rotating black hole solutions should be considered.

Motivated by the above premises, in the present paper we address the thermodynamics of Kerr-Newman black hole (KNBH) based on the nonextensive Tsallis and  R\'enyi  entropies. These black holes have been shown to reproduce a van der Waals-like equation of state under proper circumstances~\cite{Villalba}.
Special focus is reserved to thermodynamic properties and local stability in comparison to the standard Boltzmann-Gibbs framework. We also explore the nature of interactions between black hole microstructures
in the geometrothermodynamic language of Ruppeiner formalism~\cite{Rup}.

The structure of this paper is arranged as follows: in the next Section we review the four dimensional KNBH and give the conserved quantities, such as mass and angular
momentum. In Sec.~\ref{Nonext} we compute thermodynamic quantities of KNBH within Tsallis and R\'enyi statistics.
In Sec.~\ref{Stab} we investigate the thermodynamic stability
in the microcanonical and canonical treatments~\cite{Poincare}, while Sec.~\ref{Geom} is devoted to geometrothermodynamic analysis.
In last section we summarize our results and draw our conclusions.
Throughout this paper we use Planck units $c=G=\hbar=k_B=1$.
\section{Physical properties of Kerr-Newman Black Holes}
\label{KNB}
The spacetime metric that describes the $(1+3)$-dim. geometry of the Kerr-Newman (i.e. charged and rotating) black hole is identified by three parameters $(M,a,Q)$ and is presented in Boyer-Lindquist coordinates as follows~\cite{Ruiz:2019aoh}
\begin{eqnarray}
ds^2&=&\\[2mm]
\nonumber
&&\hspace{-12mm}\frac{\rho^2 \Delta}{\Sigma^2}dt^2-\frac{\rho^2}{\Sigma^2}\left(d\phi-\frac{2aMr}{\Sigma^2}dt\right)^2\sin^2\theta-
\frac{\rho^2 \Delta}{\Sigma^2}dr^2-\rho^2d \theta^2,
\label{met}
\end{eqnarray}
where
\begin{eqnarray}
\Delta&=&r^2-2Mr+a^2+Q^2\,,\\[2mm]
\rho^2&=&r^2+a^2\cos^2\theta\,,\\[2mm]
\Sigma^2&=&\left(r^2+a^2\right)^2-a^2\Delta\sin^2\theta\,.
\end{eqnarray}
In the above relations, $M$, $Q$ and $a=J/M$ are the mass,
electrical charge and specific angular momentum of the black hole,
$J$ being the total angular momentum, while $r,\theta$ and $\phi$
denote the radial and angular coordinates, respectively.

The metric~\eqref{met} has coordinate singularities in the axis of
symmetry $\theta=0$ and for $r$ such that $\Delta=0$. The latter
condition yields \be \label{eh} r_{\pm}= M\pm\sqrt{M^2-a^2-Q^2}\,,
\ee corresponding to the inner ($r_-$) and outer ($r_+$) horizons
of the KNBH metric, respectively. This relation allows us to
distinguish in principle three regimes, depending on whether
$M^2<a^2+Q^2$, $M^2>a^2+Q^2$ or $M^2=a^2+Q^2$. In what follows, we
shall consider the latter two cases, so that the square root takes
real values. Additionally, the area of the $r=r_+$ hypersurface
is~\cite{Ruiz:2019aoh}
\begin{eqnarray}
\nonumber
A&=&4\pi\left(r_+^2+a^2\right)\\[2mm]
&=&4\pi\left[2M\left(M+\sqrt{M^2-a^2-Q^2}\right)-Q^2\right].
\end{eqnarray}
Thermodynamic analysis is associated with the definition of a
thermodynamic horizon entropy. In the traditional (nonextensive)
Bekenstein-Hawking scenario, this is given by the area law
\begin{eqnarray}
\label{SBH}
S_{\rm B}=\frac{A}{4}={2\pi}\left(M^2-\frac{Q^2}{2}+\sqrt{M^4-J^2-Q^2M^2}\right),
\end{eqnarray}
which correctly recovers the Schwarzschild limit $S_B=4\pi M^2$
for $J,Q\rightarrow0$.

In its essence, the first law of black hole thermodynamics states
that the total energy in a state of thermodynamic equilibrium is
conserved. Equipped with the definition~\eqref{SBH} of entropy,
this law applied to the horizon surface takes the
form~\cite{Davies}
\begin{equation}\label{FL}
dM=TdS+\Omega dJ+\Phi dQ\,,
\end{equation}
where $T$, $\Omega$ and $\Phi$ denote the horizon temperature, angular velocity and electrical potential, respectively. On the other hand, the Smarr formula reads~\cite{Davies}
\begin{equation}
M(S,J,Q)=\left[\frac{S_B}{4\pi}\left(1+\frac{\pi Q^2}{S_B}\right)^2+\frac{\pi J^2}{S_B}\right]^{\frac{1}{2}}.
\end{equation}

One can use the first law of the thermodynamics to get the Hawking temperature, angular velocity and electrical potential as~\cite{Davies}
\begin{eqnarray}
T&=&\frac{\partial M}{\partial S_B}\,=\,\frac{1}{8\pi M}\left[1-\frac{\pi^2\left(4J^2+Q^4\right)}{S_B^2}\right],\label{T}\\[2mm]
\Omega&=&\frac{\partial M}{\partial J}\,=\,\frac{\pi J}{MS_B}\,,\label{Omega}\\[2mm]
\Phi&=&\frac{\partial M}{\partial Q}\,=\,\frac{Q}{2M}\left(1+\frac{\pi Q^2}{S_B}\right),
\label{phi}
\end{eqnarray}
respectively. Equations~\eqref{FL}-\eqref{phi} provide the basis
of black hole thermodynamics and constitute the main tools of our
next analysis.
\section{Thermodynamics of Kerr Newman BH with nonextensive entropy }
\label{Nonext}
In the previous section, we have outlined a summary of KNBH thermodynamics in the standard Bekenstein-Hawking entropy mode.
Now we examine the thermodynamic properties of the KNBH
in the case that we replace the Bekenstein-Hawking entropy with
Tsallis and R\'enyi entropies.
\subsection{Tsallis entropy}
\label{TsEntropia}

For the case of KNBH, by using Eq.~\eqref{ST1}, one can obtain the
Tsallis entropy as \be\label{ST2}
S_{T}(M,J,Q)=\gamma(2\pi)^\delta\hspace{-1mm}\left(M^2-\frac{Q^2}{2}+\sqrt{M^4-J^2-Q^2M^2}\right)^\delta,
\ee which reduces to Eq.~\eqref{SBH} for
$\delta,\gamma\rightarrow1$, as expected.

By performing straightforward algebraic manipulations,
a generalized Smarr formula for KNBH with Tsallis entropy can be obtained as
\be
\label{MT}
M^2(S,J,Q)=\frac{S_T^{\frac{1}{\delta}}}{4\pi\gamma^{\frac{1}{\delta}}}\left(1+\frac{\pi\gamma^{\frac{1}{\delta}} Q^2}{S_T^{\frac{1}{\delta}}}\right)^2+\frac{\pi\gamma^{\frac{1}{\delta}} J^2}{S_T^{\frac{1}{\delta}}}.
\ee

\begin{figure}[t]
    \begin{center}
        \includegraphics[width=8.5cm]{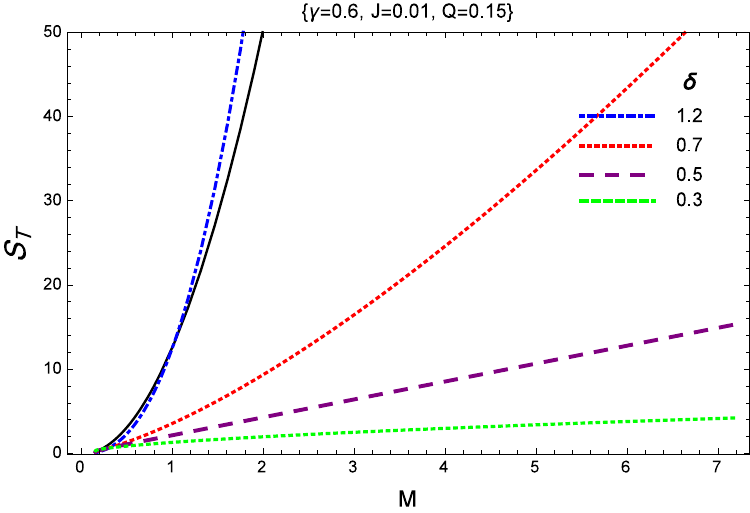}
        \includegraphics[width=8.5cm]{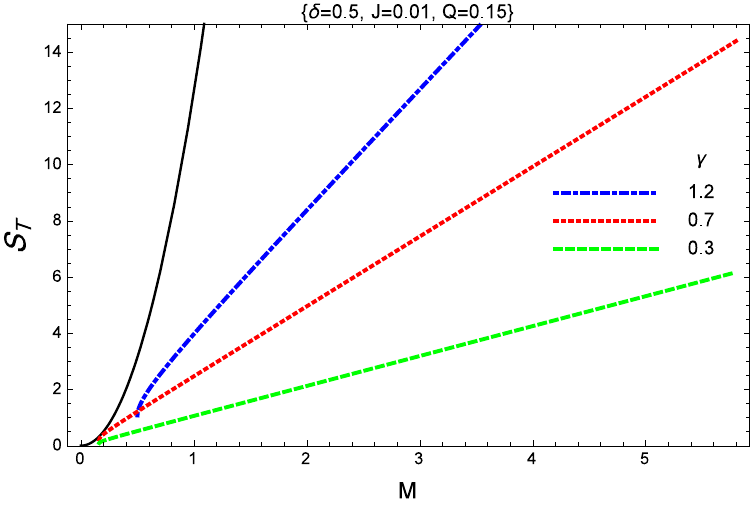}
        \caption{The Tsallis entropy $S_T$ as a function of $ M $
        for different values of $ \delta$ (upper panel)  and $
        \gamma$ (lower panel). The values $J=0.01$
        and $Q=0.15$ are adopted. The solid black curves
        correspond to the standard KNBH ($\delta=1 $
        and $ \gamma=1 $).}
        \label{figST}
    \end{center}
\end{figure}

The behavior of Tsallis entropy of the KNBH as a function of the mass is plotted in Fig.~\ref{figST}. To investigate the effect of nonextensivity, in all diagrams of the thermodynamics parameters we consider $ J $ and $ q $ as fixed and $ \delta $ and $ \gamma $ to be variable.
From  Fig.~\ref{figST} we see that both the Boltzmann (solid black curve) and Tsallis entropies of a
KNBH are monotonically increasing functions of the mass parameter.
Furthermore, the slope of the entropy curve increases
as the $\delta$ ($\gamma$) value increases.
For $ \delta>0.5 $, Kerr-Newman entropy based on the Tsallis approach becomes asymptotically convex,  similar to the  Boltzmann entropy, while it is always concave for $\delta\le0.5$. We shall see that such a different behavior has interesting implications on the classical stability analysis.

Now, by using the first law thermodynamics, we can derive the
modified Hawking temperature, angular velocity and electric
potential for KNBH as
\begin{eqnarray}
\label{TT1}
T_T(S,J,Q)&\hspace{-0.7mm}=\hspace{-0.7mm}&\frac{S_T^{\frac{1}{\delta}-1}}{8\pi\delta\gamma^{\frac{1}{\delta}} M}\left[1-\frac{4\pi^2\gamma^{\frac{2}{\delta}}}{S_T^{\frac{2}{\delta}}}\left(J^2+\frac{Q^4}{4}\right)\right],\\[2mm]
\label{OmegB}
\Omega_T(S,J,Q)&\hspace{-0.7mm}=\hspace{-0.7mm}&\frac{\pi\gamma^{\frac{1}{\delta}} J}{MS_T^{\frac{1}{\delta}}}\,,\\[2mm]
\label{PhiB}
\Phi_T(S,J,Q)&\hspace{-0.7mm}=\hspace{-0.7mm}&\frac{Q}{2M}\left(1+\frac{\pi\gamma^{\frac{1}{\delta}} Q^2}{S_T^{\frac{1}{\delta}}}\right),
\end{eqnarray}
which are in accordance with the results obtained for
Bekenestien-Hawking entropy~\cite{Davies} in the appropriate limit
$ \delta,\gamma\rightarrow1$ mentioned in the previous section.

\begin{figure}[t]
    \begin{center}
        \includegraphics[width=8.5cm]{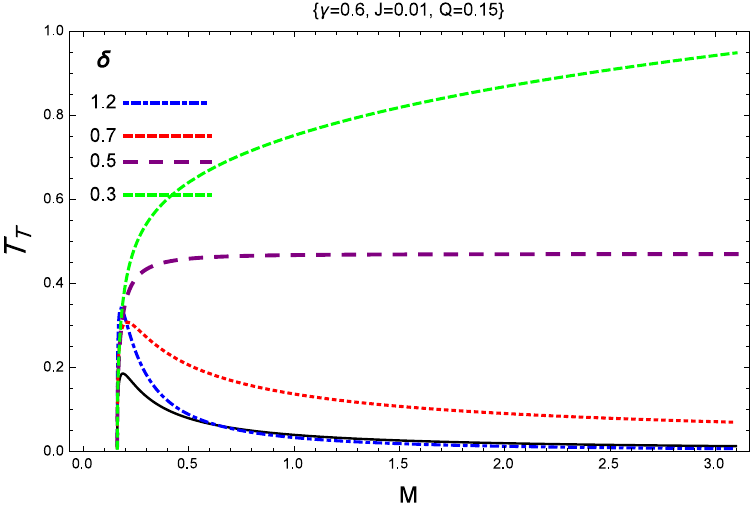}
        \includegraphics[width=8.5cm]{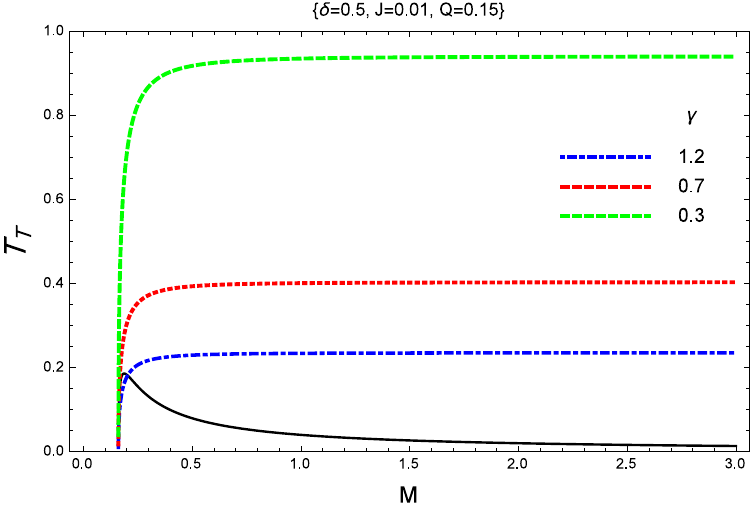}
        \caption{The Hawking temperature  $T_T$ as a function of $ M $ for Tsallis KNBH,
        for different values of $ \delta $ (upper panel) and $ \gamma $ (lower panel). The values $J=0.01$
        and $Q=0.15$ are adopted. The solid black curves correspond to the
        standard KNBH ($ \delta=1 $ and $ \gamma=1 $).}\label{figTT}
    \end{center}
\end{figure}

The behavior of the Hawking temperature is depicted in
Fig.~\ref{figTT} for various values of Tsallis parameters $ \delta
$ and $ \gamma $ and fixed $ J $ and $ Q $. According to our
former discussion on the $S$-$M$ curve, in the Tsallis approach
the temperature may have different behaviors, depending on the $
\delta $ parameter values.
As is clear from Fig.~\ref{figTT}, for $ \delta>0.5 $ (regardless of the value of $ \gamma $ parameter),
the temperature behavior of the KNBH with the Tsallis approach is similar to the standard case:
it first increases to a  
maximum point and then decreases with increasing mass.
The stationary point corresponds to the point where the heat capacity at constant $J,Q$ (which is related to the derivative of $M$ with respect to $T$) changes its sign through an infinite discontinuity (see Sec.~\ref{Stab} for more quantitative details).
On the other hand, for $ \delta<0.5 $, the temperature is a  monotonically increasing function that blows up asymptotically, while for $\delta=0.5$ it approaches  a ($\gamma$-dependent)
constant for large enough mass (see also the lower panel of Fig.~\ref{figTT}).
These features can be further checked by studying the sign of the first derivative of $T_T$ with respect to $M$ either numerically or graphically.

\subsection{R\'enyi entropy case}
\label{ReEnt}
Let us now extend the above considerations to
R\'enyi framework. By replacing the Bekenstein-Hawking entropy with the one in Eq.~(\ref{SR1}),
the KNBH entropy can be obtained as
\begin{eqnarray}\label{SR2}
\nonumber
S_{R}(M,J,Q)&=&\\[2mm]
&&\hspace{-27mm}\frac{1}{\lambda}\ln \left[1+2\pi \lambda\left(M^2-\frac{Q^2}{2}+\sqrt{M^4-J^2-Q^2M^2}\right)\right].
\label{RenEnt}
\end{eqnarray}

\begin{figure}[t]
\begin{center}
    \includegraphics[width=8.5cm]{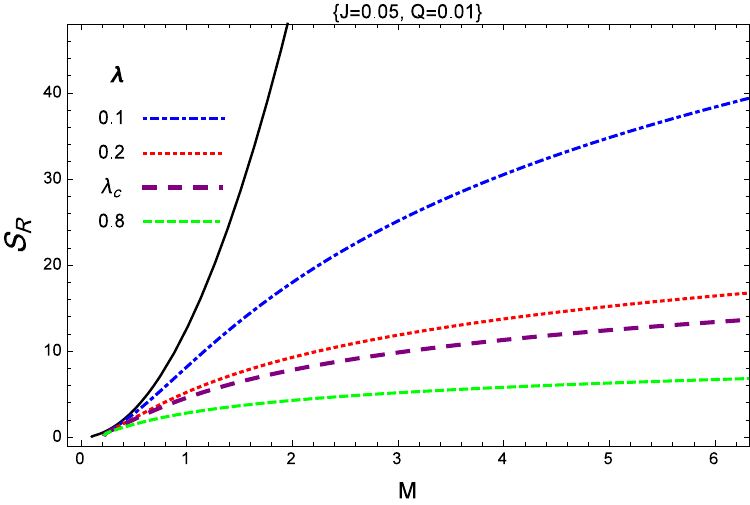}
    \caption{The entropy $S_R$ as a function of $ M $ for R\'enyi KNBH and different values of $ \lambda $. The values $J=0.05$
    and $Q=0.01$ are adopted. The solid black curve corresponds to the
    standard KNBH ($ \lambda=0 $).}
    \label{figSR}
\end{center}
\end{figure}

The behavior of R\'enyi entropy versus the mass is plotted in Fig.~\ref{figSR}, which exhibits a logarithmic growth for the selected values of $0<\lambda<1$.

For what concerns the Smarr formula, we get
\be
\label{MR}
M_{R}(S,J,Q)=\left[\frac{e^{\lambda S_R}-1}{4\pi\lambda}\left(1+\frac{\pi\lambda Q^2}{e^{\lambda S_R}-1}\right)^2+\frac{\pi\lambda J^2}{e^{\lambda S_R}-1}\right]^\frac{1}{2}\,.
\ee
Once again, by using the fist law of thermodynamics, we can calculate the thermodynamics quantities $(T, \Omega, \phi) $ as
\begin{eqnarray}\label{TR}
T_R&=&\frac{e^{\lambda S}}{8\pi M}\left[1-\frac{\pi^2\lambda^2(Q^4+4J^2)}{\left(e^{\lambda S}-1\right)^2}\right],\\[2mm]
\label{OmegaR}
\Omega_R&=&\frac{\pi\lambda J}{M\left(e^{\lambda S}-1\right)}\,,\\[2mm]
\label{phiR}
\Phi_R&=&\frac{Q}{2M}\left(1+\frac{\pi\lambda Q^2}{e^{\lambda S}-1}\right).
\end{eqnarray}

\begin{figure}[t]
\begin{center}
        \includegraphics[width=8.5cm]{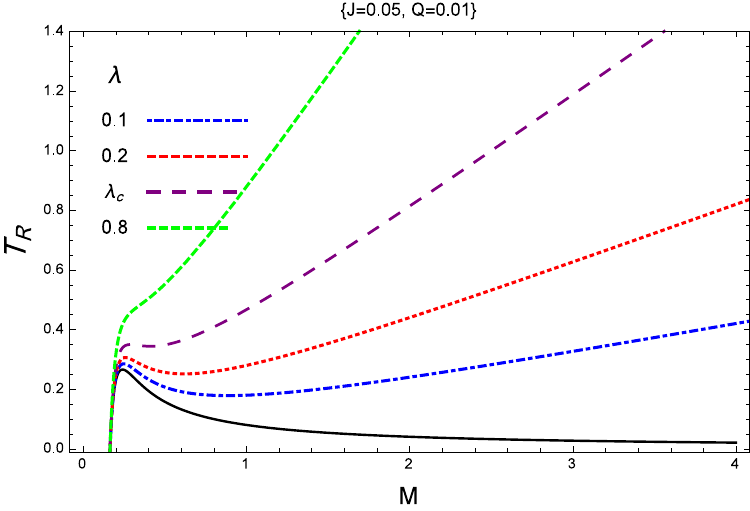}
        \caption{The Hawking temperature  $T_R$ as a function of $ M $ for R\'enyi KNBH and different values of $ \lambda $. The values $J=0.05$
            and $Q=0.01$ are adopted. The solid black curve corresponds to the
            standard KNBH ($ \lambda=0 $).}\label{figTR}
\end{center}
\end{figure}

We have plotted the behavior of the temperature for KNBH based on R\'enyi entropy in Fig.~\ref{figTR} for different values of $ \lambda $ parameter. As it is clear, for $ \lambda=0 $ the standard case is reproduced and the temperature curve exhibits a global maximum.
On other hand, in the generalized case based on the R\'enyi entropy and for the selected values of model parameters and $0<\lambda<\lambda_c\simeq0.39$ 
the temperature curves have a local maximum and a local minimum, which indicates that there are three distinct regimes of black holes:
small black hole (where $\partial_M T$ is positive), intermediate black hole (where $\partial_M T$ takes negative values) and large black hole (where $\partial_M T$ turns out to be positive again).
Such a behavior reveals that, as the mass increases, the black hole undergoes a small-large black hole transition, which can be described by the language of first order phase transformations in Van der Waals fluids~\cite{Villalba} (see also the next stability analaysis). The two stationary points correspond to the points where the heat capacity at constant $J,Q$ changes its sign through an infinite discontinuity, similar to the Davies point of the standard KBH case~\cite{Davies}.
Similar features have been achieved in~\cite{Luciano} and~\cite{Luciano:2023bai} for the case of Tsallis and Kaniadakis entropies in charged AdS black holes, respectively.

As the nonextensive parameter increases at the critical value
$\lambda=\lambda_c $,
the local maximum and minimum converge into a single inflection point, which means that the intermediate region becomes gradually thinner and finally collapses into a point. This can be checked by computing the second derivative of $T_R$ with respect to $M$ and verifying that it vanishes in this point. Such a behavior signals
that the transition becomes second order (see also the pertinent discussion in Sec.~\ref{Stab}).
Above $\lambda_c$, the intermediate region disappears
and the temperature curve is monotonically increasing function of the mass parameter.

\section{Stability analysis}
\label{Stab}
The thermal stability of black holes is a crucial aspect to consider
when studying their thermodynamic properties.
For this purpose, we analyze the stability of the
KNBH solutions against axially symmetric perturbations
by taking into account the effects of entropy's
nonextensivity property in both the microcanonical and canonical ensemble.
Initially, we examine the pure case of isolated black hole using the microcanonical ensemble.
We then shift our focus to the canonical ensemble, wherein the black holes are in contact with a surrounding thermal radiation bath.

In ordinary thermodynamics of extensive systems, the local thermodynamic stability is linked to the dynamical stability of
the system and is investigated by Hessian method, which consists in analyzing the behavior of the heat capacity.
However, this analysis strongly depends on the
additive property of the entropy function and, for this reason, its application to black holes in the conventional Boltzmann framework may
lead to tricky results~\cite{Czinner4}. The most glaring example is the case of the Schwarzschild black hole, which is known to be perturbatively stable, although the Hessian method indicates the opposite behavior. On the other hand, this method has been recently applied to study the phase transitions of AdS black holes in Tsallis~\cite{Morad1,Luciano}, Barrow~\cite{Jawad} and Kaniadakis~\cite{Luciano:2023bai} entropy. The obtained results align with the
known thermodynamic properties of standard AdS black holes and show no apparent contradiction.

To avoid any possible pitfall arising from the application of the Hessian
approach to the stability analysis, a different strategy has been proposed by Kaburaki et al.~\cite{Kaburaki1,Kaburaki1bis}. In this study, the authors make use of the Poincar\'e turning point method~\cite{Poincare} to investigate the thermodynamic stability of black holes.
This method is based on topology and does not rely on the additivity of the entropy function.
Later on, it has been employed to examine the stability of several classes of black holes~\cite{Davies,Kaburaki1,Kaburaki1bis,Kaburaki2,Czinner,Czinner2,Czinner3,Czinner4} and in particular of Schwarzschild black holes with R\'enyi entropy~\cite{Czinner2}. In this case, a Hawking-Page phase transition has been shown to occur in a similar fashion as in AdS space in the Boltzmann statistics, the corresponding critical temperature depending only on the deformation entropic parameter. Additionally, the stability analysis has been conducted with both the Poincar\'e and the Hessian methods. The latter could be applied since R\'enyi
entropy is additive for composition.

To understand the Poincar\'e method,
we briefly review it by following~\cite{Kaburaki1,Kaburaki1bis,Kaburaki2,Czinner,Czinner2,Czinner3,Czinner4}. The proofs and details are given in the original references.
Let us suppose there is a distribution function called $ Z(x^i,y) $, whose extremal points are defined by
$ \frac{\partial Z}{\partial x^i}=0 $.
When the extremal of $ Z $ is a maximum,
it signifies stable equilibrium configurations.
Now, let us consider the equilibrium $ Z(y)=Z[X^i(y),y]$,
where $ X^i(y) $ is a solution of
$ \frac{\partial Z}{\partial x^i}=0 $.
If the derivative function $ dZ/dy $, which is considered against $ y $,
forms a continuous and differentiable curve, changes in stability only occur at points where the tangents are vertical.
The $ Z $ distribution function is known as the Massieu function,
and the points with vertical tangents are referred to as turning points.
The regions near the turning points with negative tangents represent a
branch of unstable configurations, the regions with positive slopes near
the turning points signify a branch of more stable configurations.

To ensure comprehensive coverage, in the next analysis of thermodynamic stability we examine both the microcanonical and the canonical ensembles. {Since R\'enyi entropy is additive for composition~\cite{Czinner4}, stability of R\'enyi KNBH will be investigated by using both the Poincar\'e and Hessian methods, showing consistency of ensuing results. The same comparative analysis is also conducted for the case ot Tsallis KNBH. In spite of non-additivity of Tsallis entropy~\cite{Tsallis}, we find that Poincar\'e and Hessian approaches still predict the same thermodynamic properties. This substantiates the view that
thermodynamics of non-extensive systems might hide pitfalls and one has to be very careful when considering stability and phase transitions of such complex systems.}

\subsection{Microcanonical ensemble}

\begin{figure}[t]
\begin{center}
        \includegraphics[width=8.5cm]{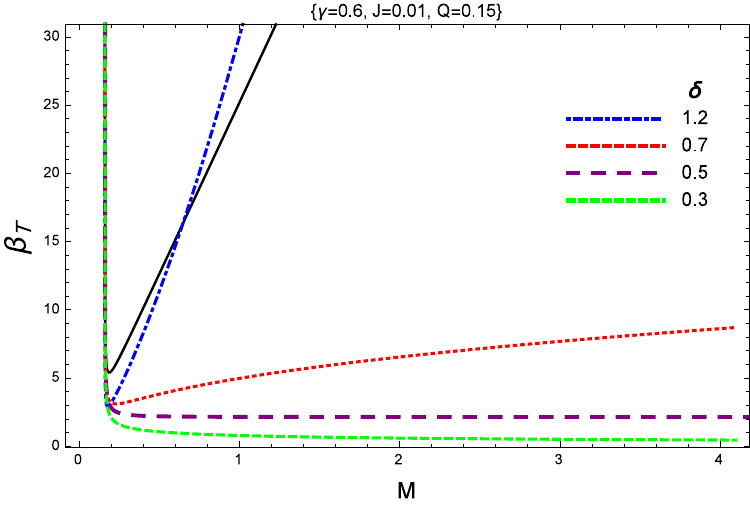}
        \includegraphics[width=8.5cm]{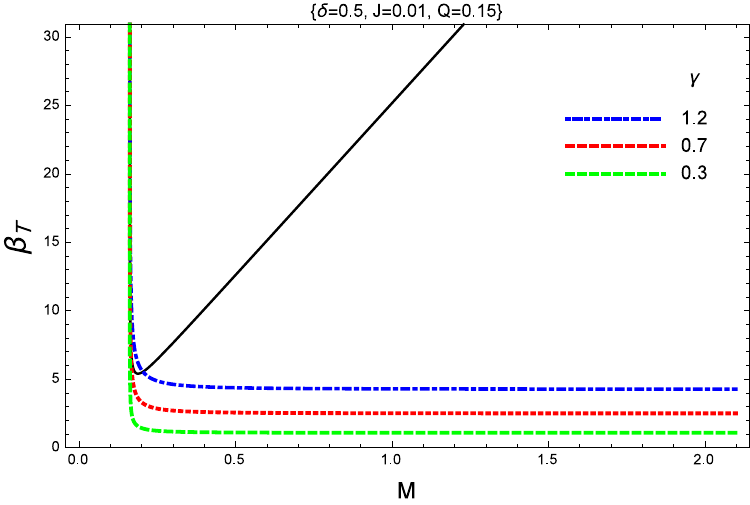}
        \caption{The conjugate parameter  $\beta_T(M)$ for Tsallis
        KNBH and different values of $ \delta $ (upper panel) and $
        \gamma $ (lower panel). The values
        $J=0.01$ and $Q=0.15$ are adopted. The solid black
        curves correspond to the standard KNBH ($
        \delta=1 $ and $ \gamma=1 $).}
        \label{figBT}
\end{center}
\end{figure}

\begin{figure}[t]
\begin{center}
        \includegraphics[width=8.5cm]{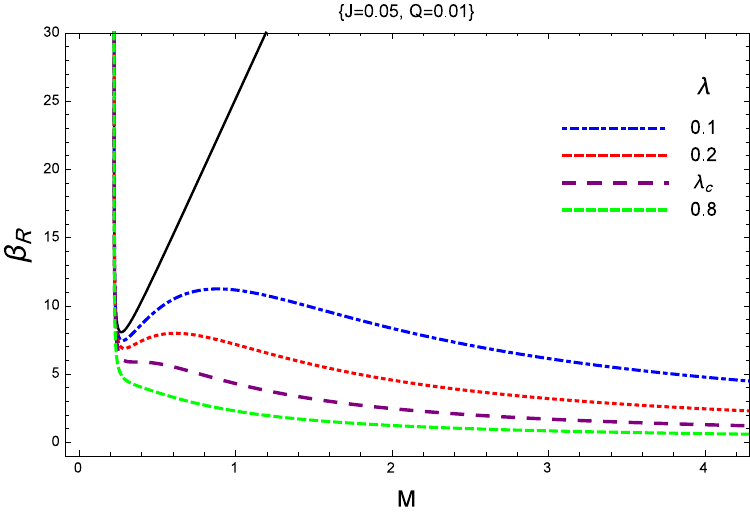}
        \caption{The conjugate parameter $\beta_R(M)$ for R\'enyi
        KNBH,
        for different values of $ \lambda $ and fixed
        $J=0.05$ and $Q=0.01$ are adopted. The solid black curve corresponds to the
        standard KNBH ($ \lambda=0 $).}\label{figBR}
\end{center}
\end{figure}

\begin{figure}[t]
\begin{center}
        \includegraphics[width=8.5cm]{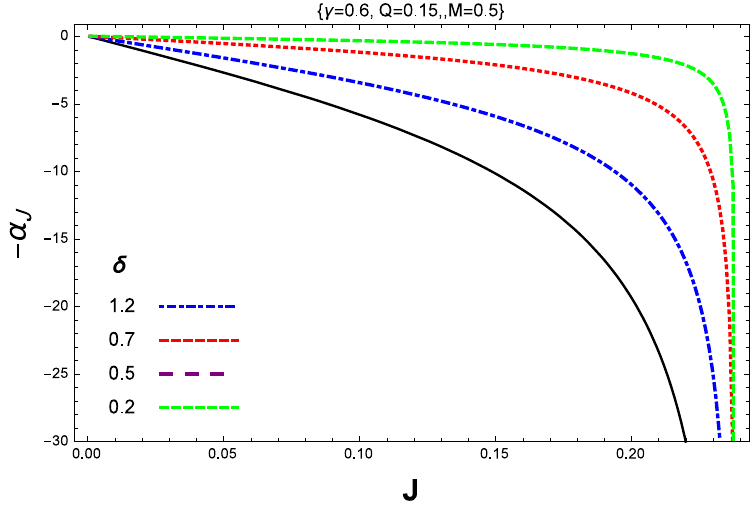}
        \includegraphics[width=8.5cm]{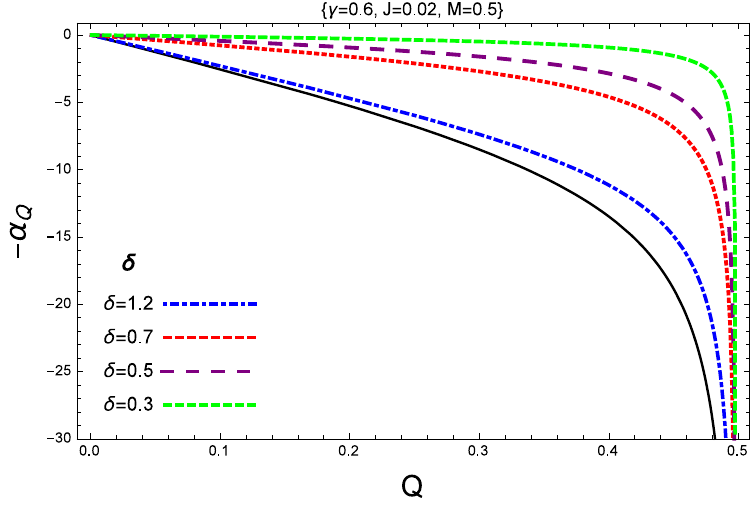}
        \caption{The conjugate parameter $\alpha_J(J)$ and $ \alpha_Q(Q) $ for Tssalis KNBH at constant $ \gamma $ and $ M $, and for different values of $ \delta $.
        The solid black curves correspond to the standard KNBH ($ \delta=1 $ and $ \gamma=1 $).}
        \label{figaT}
\end{center}
\end{figure}

\begin{figure}[t]
\begin{center}
    \includegraphics[width=8.5cm]{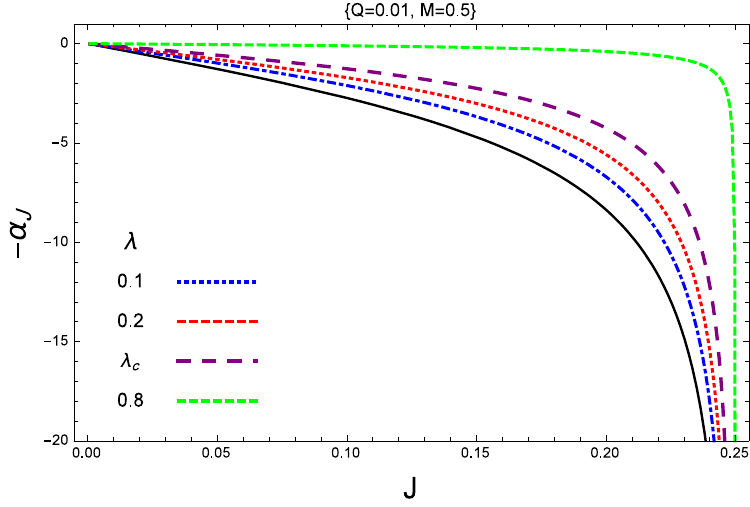}
    \includegraphics[width=8.5cm]{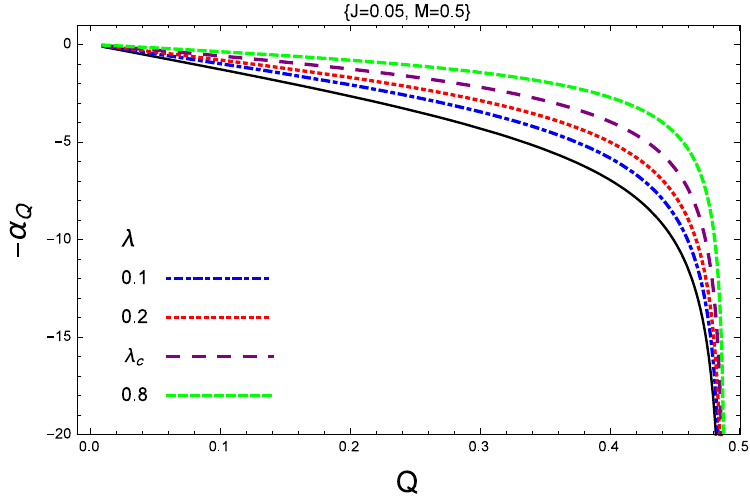}
    \caption{The conjugate parameter $\alpha_J(J)$ and $ \alpha_Q(Q) $ for R\'enyi KNBH at constant $ M $ and different values of $ \lambda $.
    The solid black curves correspond to the standard KNBH ($ \lambda=0 $).}\label{figaR}
    \end{center}
\end{figure}

As mentioned above,  according to Poincar\'e turning point method~\cite{Poincare}, change of the stability in an equilibrium state (stable or not)
only occurs at turning points which are extremal of the Massieu function.
For KNBH based on the Tsallis and R\'enyi statistics, this function is given by the entropies~(\ref{ST2}) and~(\ref{SR2}),
respectively, where $ M $, $ J $ and $ Q $ act as control parameters.
Each of these parameters has a conjugate
counterpart  defined by
\be
\label{dsNEW}
ds=\beta dM-\alpha_J dJ-\alpha_Q dQ\,,
\ee
where
\be
\beta=\frac{1}{T}\,,\quad \alpha_J=\frac{\Omega}{T}\,, \quad  \alpha_Q=\frac{\Phi}{T}\,.
\ee



To perform the stability analysis in the present case, we have to plot these conjugate parameters. Fig.~\ref{figBT} and
Fig.~\ref{figBR} display the behavior of $\beta(M)$ for KNBH based on Tsallis and R\'enyi entropy, respectively.
As is evident, there is no vertical tangent point, similar to the standard KNBH (solid black curve)~\cite{Kaburaki2}. This means that there is no stability change for any $M$ in the microcaninical ensemble.
Such a feature is confirmed by further investigation of the behavior of $ -\alpha_J(J) $ and  $ -\alpha_Q(Q)$ in Fig.~\ref{figaT} and Fig.~\ref{figaR}.
Therefore, we can conclude that isolated KNBH based on
Tsallis and R\'enyi entropies are locally stable concerning axially symmetric perturbations. This result is in accordance with~\cite{Czinner3}.

From Fig.~\ref{figBT} we can also see that, for $\delta>0.5$, there is
a point where the tangent of the stability curves becomes horizontal.
This corresponds to the point where the heat capacity at constant
$J,Q$ changes its sign through an infinite discontinuity, see also the discussion at the end of Sec.~\ref{TsEntropia}. Similarly, from Fig.~\ref{figBR} we infer that there are two points where the tangent of the stability curves is horizontal as far as $0<\lambda<\lambda_c$, while no one occurs for $\lambda>\lambda_c$. This aligns with the considerations at the end of Sec.~\ref{ReEnt} and results in~\cite{Czinner3}.
\subsection{Canonical ensemble}
In the following, we investigate the thermodynamic stability of
the black hole in the canonical approach using both Poincar\'e and Hessian methods.
Since we consider the model based on the additive the Tsallis and R\'enyi entropies,
and hence the standard Hessian approach is also applicable.

The canonical ensemble is used to describe the state of a black hole that is in equilibrium with an infinite reservoir of thermal radiation at fixed temperature.
The Massieu function in this case is $Z=S-\beta M=-\beta F$,
where
\be
\label{FEn}
F=M-TS
\ee
is the Helmholtz free energy
and the conjugate variables of the control parameters are $ -M $, $ -\alpha_J $ and $ -\alpha_J $.
To study the stability in this case, we need to plot the
curves $ -M(\beta) $  at constant $ J $ and $ Q $, and  $ J(\alpha_J) $ and  $ Q(\alpha_Q) $ at constant $ \beta $.

In fact, the stability curves of $-M(\beta)$ at constant $J$ for Tsallis and R\'enyi models are essentially the $\frac{\pi}{2}$ clockwise rotated versions of the Figs.~\ref{figBT} and~\ref{figBR}, respectively.
For the KNBH based on the Tsallis approach, looking at the rotated Fig.~\ref{figBT},
we can see that there is a vertical tangent (turning point) along the stability curve for
$ \delta>0.5 $ (independent of the $ \gamma $ value), where the stability of system changes.
The branch with positive slope indicates a class of small black holes that are more stable than large black holes with negative slope.
The standard Boltzmann framework is recovered by continuously
varying $\delta$ to unity.
On the other hand, for  $ \delta\leqslant 0.5 $, the stability curve does not have any vertical tangent point and is always positive.

For the R\'enyi approach, examining the rotated Fig~\ref{figBR},
we infer that there is a vertical tangent along the stability
curve in the standard Boltzmann treatment $\lambda=0$. On the
other hand, for $0<\lambda<\lambda_c$ we see that there are two
vertical tangent points, which implies that the thermodynamic
stability undergoes two changes. Between these two turning points,
we have three different classes of black holes: a small and large
stable black holes (with a positive curve slope), and an unstable
intermediate one (with a negative curve slope). Therefore, one can
expect a first order small black hole/large black hole phase
transition in a very similar fashion as in AdS space. As $\lambda$
increases to the critical value $\lambda_c$, these two points
collapse into a single one, corresponding to the critical
inflection point mentioned at the end of Sec.~\ref{ReEnt}. For
$\lambda>\lambda_c$ no vertical tangent point occurs. In both
cases, there is no stability change. These findings are consistent
with~\cite{Czinner3}.

{To thoroughly explore stability in the canonical ensemble,
we also apply the traditional Hessian method to Tsallis and R\'enyi  framework.}
Since the angular momentum and charge parameters are considered constant, the only sign of heat capacity is sufficient to ensure stability.
In this way, for $ C_{J,Q}>0 $ the black hole system is stable, while $ C_{J,Q}<0 $ means instability.

The heat capacity of the KNBH is defined as
\be
\label{C}
C_{J,Q}=T\left(\frac{\partial S}{\partial T}\right)_{J,Q}\,.
\ee
{From Eq.~(\ref{TT1}), we get
\begin{widetext}
\begin{eqnarray}\label{C_T}
C_{J,Q}\left(M\right)&=&\left[2^{3+\delta}M^2\pi^{2\delta}f_{J,Q}\left(2M^2-Q^2+2f_{J,Q}
\right)^{\delta+1}\gamma\delta
\right]\Bigg\{\left(2\pi\right)^\delta f_{J,Q}\left[Q^4-\left(2M^2-Q^2+2f_{J,Q}
\right)^2
\right]\\
\nonumber
&& +\,2^{2+\delta}\pi^\delta J^2\left[2M^2\left(1+\delta\right)+f_{J,Q}\right]-2^{1+\delta}\pi^\delta M^2\left[\left(2M^2-Q^2+2f_{J,Q}\right)^2(\delta-1)-Q^4\left(1+\delta\right)
\right]
\Bigg\}^{-1}
\end{eqnarray}
\end{widetext}
where we have introduced the simplified notation $f_{J,Q}(M)\equiv\sqrt{M^4-J^2-Q^2M^2}$. The above expression
recovers the result of employing the Bekenstein
entropy at the appropriate limit $\delta=\gamma=1$, namely
\be
\label{lim}
C_{J,Q}^{\delta=1=\gamma}(M)=8\pi M^2\hspace{-0.5mm}\left\{-1+\frac{\left(4J^2+Q^4\right)\left(4M^2+f_{J,Q}\right)}{f_{J,Q}\left(Q^2-2M^2-2f_{J,Q}\right)^2}
\right\}^{-1}\,.
\ee
}

\begin{figure}[t]
\begin{center}
        \includegraphics[width=8.5cm]{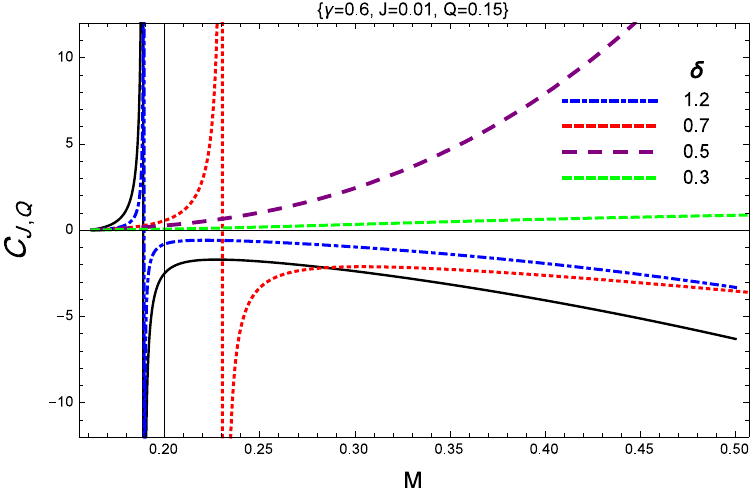}
        \includegraphics[width=8.5cm]{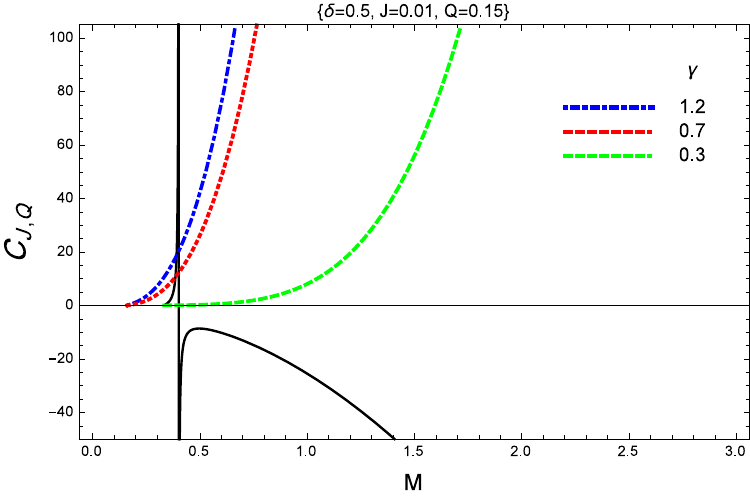}
        \caption{The heat capacity  $C_{J,Q}(M)$ for Tsallis KNBH,
        for different values of $ \delta $ and $ \gamma $
        where the values of $ J=0.01 $ and $ Q=0.15 $ are adopted.}\label{figCT}
\end{center}
\end{figure}

{We have displayed the behavior of $ C_{J,Q}$  versus $ M $ for Tsallis case in Fig.~\ref{figCT} for various values of $ \delta $ and fixed $ J $,$ Q $.
The plot demonstrates that, for $ \delta>0.5 $, the heat capacity $ (C_{J,Q}) $
diverges at a given mass point, where the temperature is maximum, and passes from positive values (more stable configuration) to negative values (less stable configuration) for increasing $M$~\cite{Davies}.
This discontinuity also corresponds to the point where the tangent of the stability curve $ -M(\beta) $
becomes vertical.
For the $ \delta\leq 0.5 $ case,  $ C_{J,Q} $ has no divergent point and it is always positive.
We notice that the obtained results are completely consistent with the Poincar\'e method.}
%

Now, concerning R\'enyi KNBH, one can use the temperature in Eq.~(\ref{TR}) to obtain
\begin{widetext}
\begin{eqnarray}
\label{CRen}
C_{J,Q}\left(M\right)&\hspace{-1mm}=\hspace{-1mm}&\left[2\pi M^2f_{J,Q}\left(Q^2-2M^2-2f_{J,Q}\right)^2
\right]\hspace{-1mm}\Bigg\{2M^2Q^2f_{J,Q}-4\pi\lambda J^4+8\pi \lambda M^8-M^6\left[2-8\pi\lambda\left(f_{J,Q}-Q^2\right)\right]+M^4\\
\nonumber
&&\hspace{-20mm}\times\left\{Q^2\left[3+\pi\lambda\left(Q^2-4f_{J,Q}\right)\right]-2f_{J,Q}\right\}+J^2\left\{2f_{J,Q}+4\pi\lambda M^4+Q^2\left[\pi\lambda\left(Q^2-4f_{J,Q}\right)-1\right]+M^2\left[6+8\pi\lambda\left(f_{J,Q}-Q^2\right)\right]
\right\}
\hspace{-1mm}\Bigg\}^{-1},
\end{eqnarray}
\end{widetext}
which consistently gives the limit in Eq.~\eqref{lim} for $\lambda\rightarrow0$.

\begin{figure}[t]
\begin{center}
        \includegraphics[width=8.3cm]{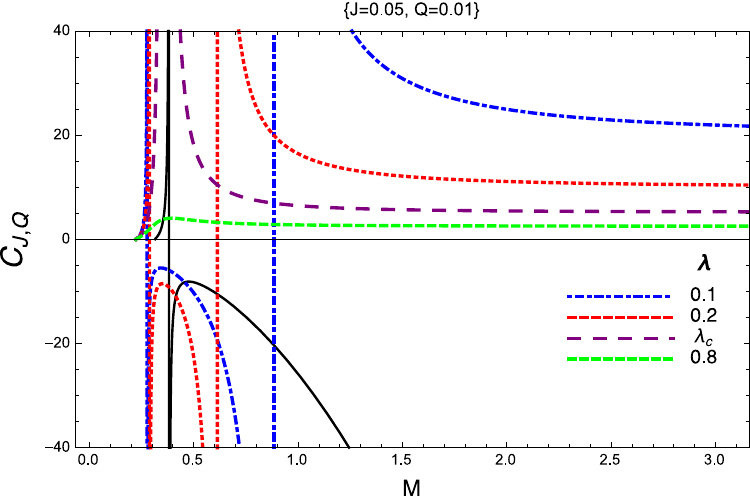}
        \caption{The heat capacity  $C_{J,Q}(M)$ for R\'enyi KNBH,
        for different values of $ \lambda $
        where the values of $ J=0.05 $ and $ Q=0.01 $ are adopted.}\label{figCR}
\end{center}
\end{figure}

The behavior of $C_{J,Q}$ in Eq.~\eqref{CRen} has been illustrated in Fig.~\ref{figCR} for various values of $\lambda$.
For the standard Boltzmann case $(\lambda=0)$, there is a value of $M$ such that $C_{J,Q}\rightarrow\infty$ and changes its sign passing from positive to negative values as $M$ increases. This occurrence corresponds to the turning point in the Poincar\'e method.
In the R\'enyi approach, the heat capacity curve strongly depends on the value of the $\lambda$ parameter.
For $0<\lambda<\lambda_c$, the heat capacity curve has two divergences,
which correspond to the two turning points. As a result,
we can distinguish three configurations of black holes, in line with the discussion below Eq.~\eqref{phiR}:
small and large black holes are found within the region where the heat capacity is positive and show stable behavior, while
intermediate black holes have negative heat capacity
and are unstable. In this case, we have a first order transition between the small and large branches.
At $\lambda=\lambda_c$, $C_{J,Q}$ exhibits
one divergence but no change of sign around it, which signals the occurrence of the critical (inflection) point where the phase transition
becomes second order.
Finally, for $ \lambda> \lambda_c $, the heat capacity is always continuous (i.e., it has no turning points in the Poincar\'e method language) and positive, which entails stable behavior at any $M$.

\subsubsection{Phase transitions of R\'enyi Kerr-Newman black holes}
\begin{figure}[t]
\begin{center}
        \includegraphics[width=8.3cm]{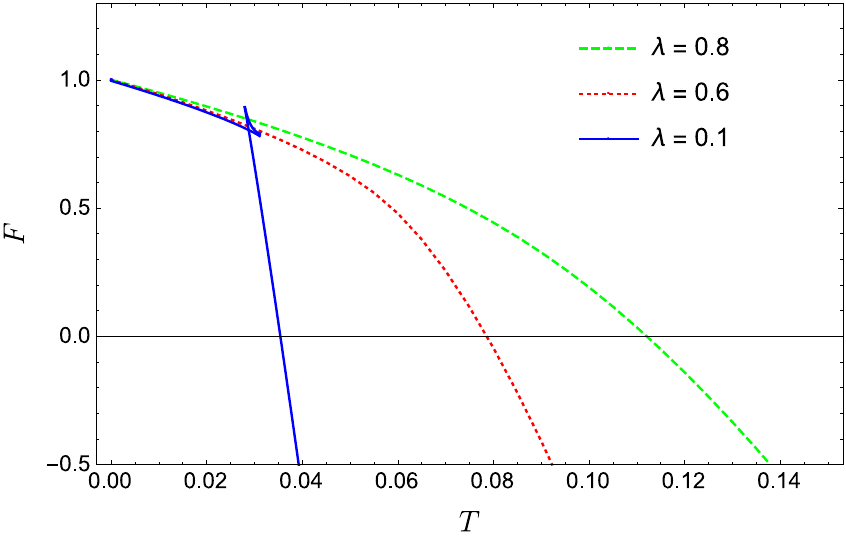}
        \caption{The free energy against temperature for R\'enyi
        KNBH and different values of $\lambda$. The
        characteristic swallow-tail behavior is observed below the
        critical value of $\lambda$, which corresponds to a first order
        small black hole/large black hole transition.}
        \label{FigFR}
\end{center}
\end{figure}

In this subsection we pursue the matter of possible phase
transitions of KNBH. We focus on the case of R\'enyi KNBH, which
have been shown to exhibit a rich thermodynamic phase structure.
Valuable information in the characterization of black holes phase
transitions is provided by the study of free energy~\eqref{FEn}.
The behavior of $F$ versus $T$ is displayed in Fig.~\ref{FigFR}.
In compliance with our former discussion, we observe that all
curves cross the horizontal ($F=0$) axis. In particular, small
KNBH with lower temperature have positive free energy, in contrast
to larger KNBH that have negative values of $F$. This feature is
indicative of a Hawking-Page like transition between the thermal
gas phase with angular momentum and the large black hole state,
which is locally stable (see the considerations at the end of the
previous subsection). As noted in~\cite{Czinner4}, it is typically
assumed that the free energy for a thermal gas is nearly
vanishing. Hence, the phase transition is expected to occur when
the black hole reaches a temperature at which its free energy is
also zero.

Along with the Hawking-Page phase transition, a small black
hole/large black hole transition takes place as far as the
entropic parameter $\lambda$ lies below the critical value (blue
curve in Fig.~\ref{FigFR}). Such a phenomenon manifests through
the characteristic swallow-tail behavior of free energy, which is
typical of a first order transition between the small and large
black hole phases. In line with our findings in the Poincar\'e and
Hessian methods, we here identify three different branches:  small
black holes at lower temperature, large black holes at higher
temperature and intermediate unstable black holes. Following the
blue curve as $T$ increases, we see that the KNBH first runs along
the small branch until $T$ increases to the critical point of
phase transition. At this point, the small and large phases
coexist, since they have the same free energy. For higher $T$,
large black holes become more stable and represent the favored
configuration.

For $\lambda$ above the critical value (red and green lines in
Fig.~\ref{FigFR}), the swallow-tail behavior disappears, as one
would expect from the absence of turning points in the rotated
curves of Fig.~\ref{figBR} and divergence points in the heat
capacity of Fig.~\ref{figCT}. This reveals that the phase
transition first becomes second order at the critical value
$\lambda=\lambda_c$ and then does not occur anymore for
$\lambda>\lambda_c$. Notice that all these results and
considerations agree with those in~\cite{Czinner4} for the case of
Kerr black holes.

\section{Geometrothermodynamics analysis}
\label{Geom}

It was suggested by Weinhold~\cite{Wei} and Ruppeneir~\cite{Rup}
that interesting properties of thermodynamic systems can be
understood from a geometric perspective. The strategy is to define
a Riemannian metric as the second derivative of internal
energy~\cite{Wei} or entropy~\cite{Rup} with respect to
thermodynamic variables of the system. The ensuing framework is
commonly referred to as \emph{geometrothermodynamics} and
represents an effort to infer microscopic interaction information
from the axioms of thermodynamics phenomenologically or
qualitatively. The basic idea is that the sign of the scalar
curvature $R$ of the metric can be related to the nature of
microscopic interactions. In particular, negative (positive)
curvature means prevailing attraction (repulsion), while flatness
corresponds to systems where interactions are perfectly balanced.

After the development of black hole mechanics,
geometrothermodynamic techniques have been largely applied to
black holes to address the phenomenological micro-mechanism of
sub-interactions. Studies in this direction appear
in~\cite{Cai,PRL,WeiLiu,Xu,Ghosh} and recent extensions to
non-extensive statistics have been conducted
in~\cite{Luciano,Luciano:2023bai,Jawad}. In what follows, we
resort to Ruppeneir formalism~\cite{Rup}. By considering the
system entropy as thermodynamic potential, Ruppeneir line element
is specified as \be \label{Rupme}
ds^2_R=g_{\alpha\beta}^Rdx^\alpha dx^\beta\,, \quad
g^R_{\alpha\beta}=-\partial_\alpha\partial_\beta S\,, \ee where we
have denoted the independent fluctuation coordinates by $x^i$. In
passing, we would like to observe that this geometry is related to
the Weinhold mass representation by $ds^2_R=ds^2_W/T$, where the
Weinhold line element is defined by
$ds^2_W=g_{\alpha\beta}^Wdx^\alpha dx^\beta$,
$g_{\alpha\beta}^W=-\partial_\alpha\partial_\beta M$.

As discussed in~\cite{Rup2}, one has three different settings for KNBH, depending on the thermodynamic variables that are considered  as fluctuating. We shall focus on the two cases: \emph{1)} $(M, Q)$ fluctuating, $J$ fixed and \emph{2)} $(M, J)$ fluctuating, $Q$ fixed.
The case where all $(M,J,Q)$ fluctuate (corresponding to a $n=3$ Riemannian geometry) falls entirely outside the domain of stable fluctuations. Similarly, $M,J$ and $Q$ fluctuations with $n=1$ have trivially vanishing $R$ and will not be studied in the following.

\subsection{Tsallis entropy}
By use of Eq.~\eqref{Rupme}, we can now compute Ruppeneir scalar curvature for KNBH. In the case of Tsallis entropy~\eqref{ST2}, the following results are obtained:

\subsubsection{$(M, Q)$ fluctuating, $J$ fixed}

\begin{figure}[t]
    \begin{center}
        \includegraphics[width=8.5cm]{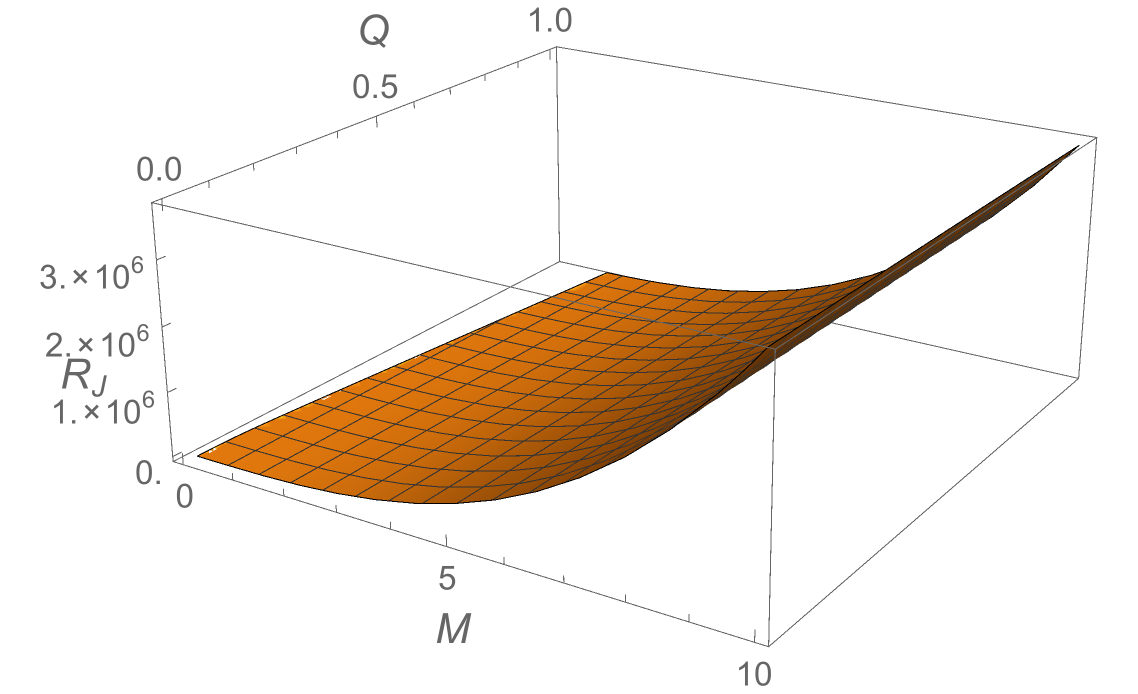}
        
        \vspace{4mm}
             \includegraphics[width=8.3cm]{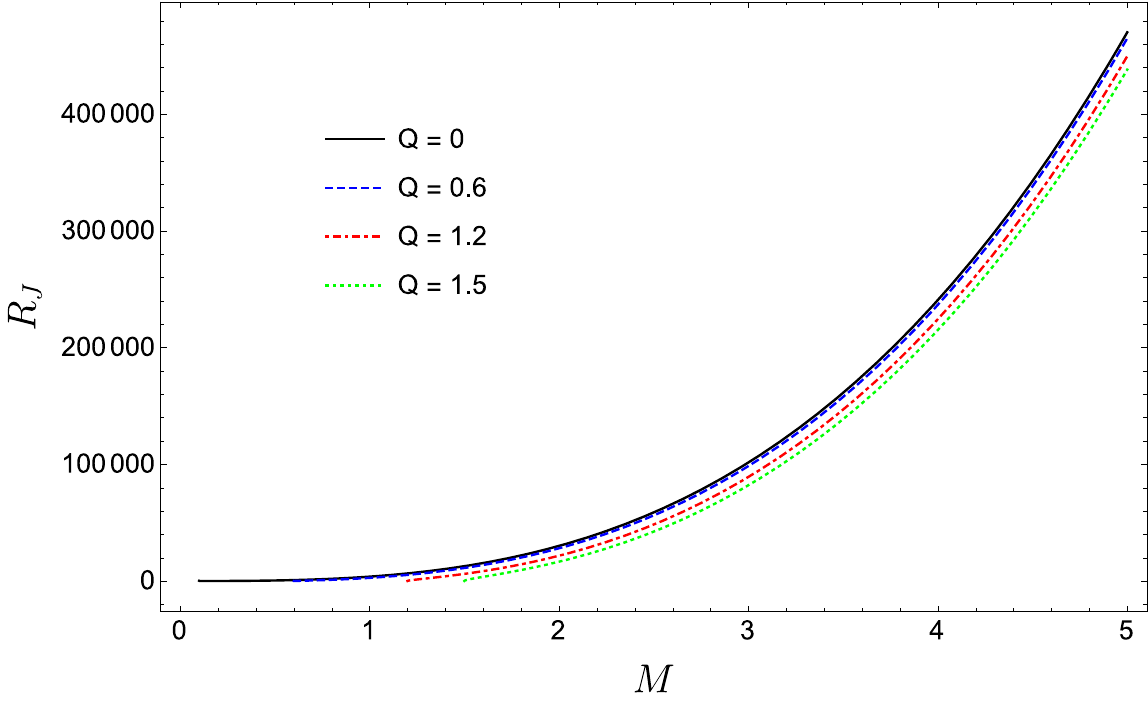}
             \vspace{3mm}
                  \includegraphics[width=8.3cm]{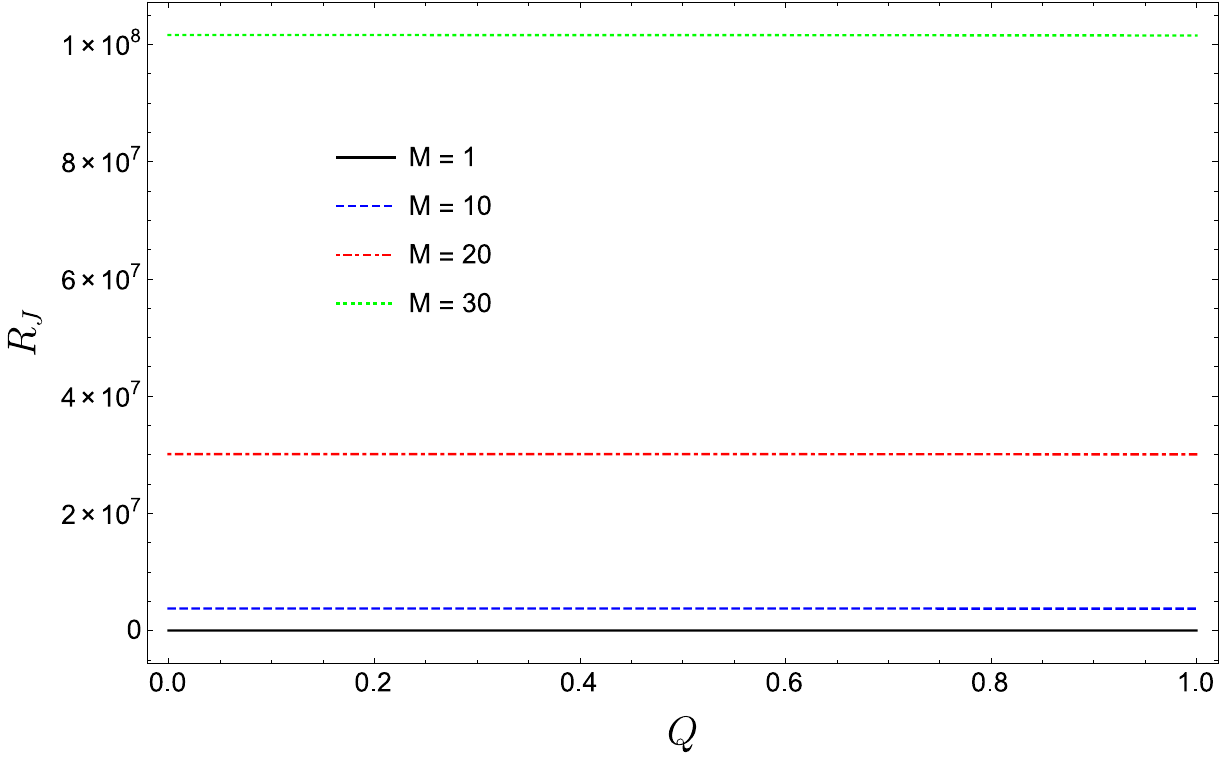}
        \caption{Ruppeneir curvature of Tsallis KNBH for fixed $J$ and fluctuating $(M,Q)$. We set $\delta=0.5$.}
        \label{Ruppe1}
    \end{center}
\end{figure}

Here, $(M, Q)$ fluctuate at fixed $J$. Accordingly, the scalar curvature can be computed by implementing $dx^J=0$ in Eq.~\eqref{Rupme}. We get
\be
\label{met1}
ds_R^2=g_{MM}\left(\Delta M\right)^2+2g_{MQ} \left(\Delta M\right) \left(\Delta Q\right)+g_{QQ}\left(\Delta Q\right)^2\,,
\ee
which corresponds to a 2-dim Riemannian geometry.
The explicit expression of the scalar curvature $R_J$ is rather awkward.
Nevertheless, for the purpose of our study, it is enough to
resort to a graphical analysis.
For the selected values of model parameters and $\delta\le0.5$, we can see that $R_J$ is real everywhere in the physical regime (see the upper panel of Fig.~\ref{Ruppe1}). Moreover, it is positive and increases for increasing black hole mass, which signifies repulsive interactions, as
it is evident from the middle panel. For fixed $M$, the curvature remains nearly constant as $Q$ varies (lower panel).

\begin{figure}[t]
    \begin{center}
        \includegraphics[width=8.5cm]{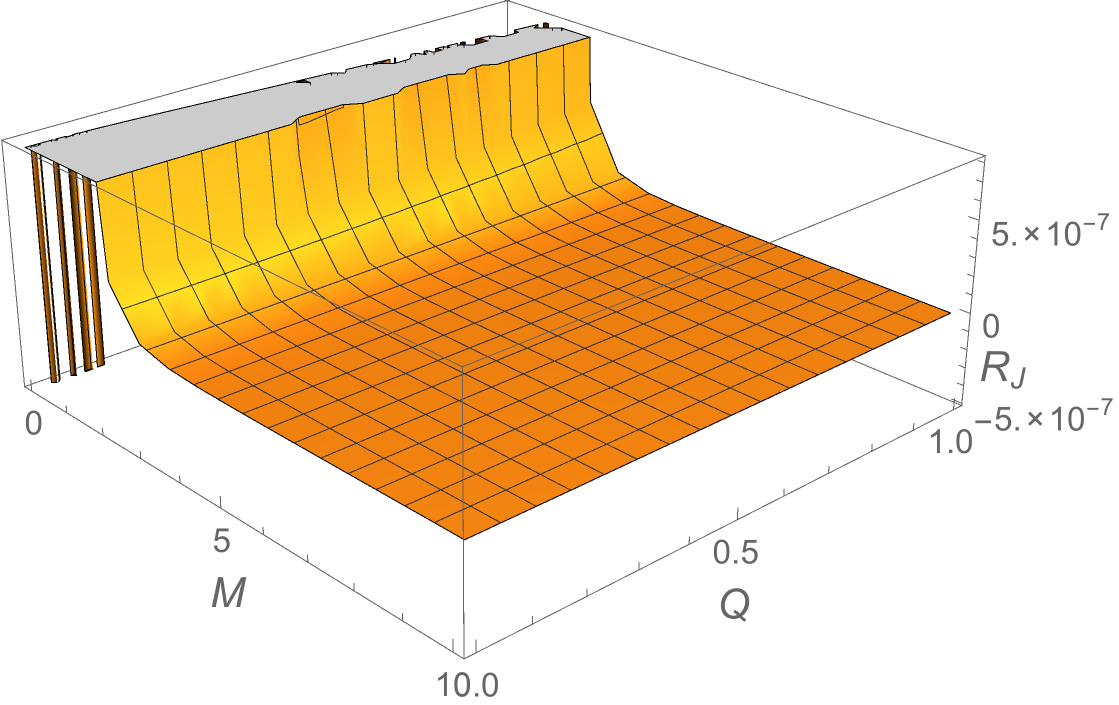}
        
              \vspace{4mm}
         \includegraphics[width=8.3cm]{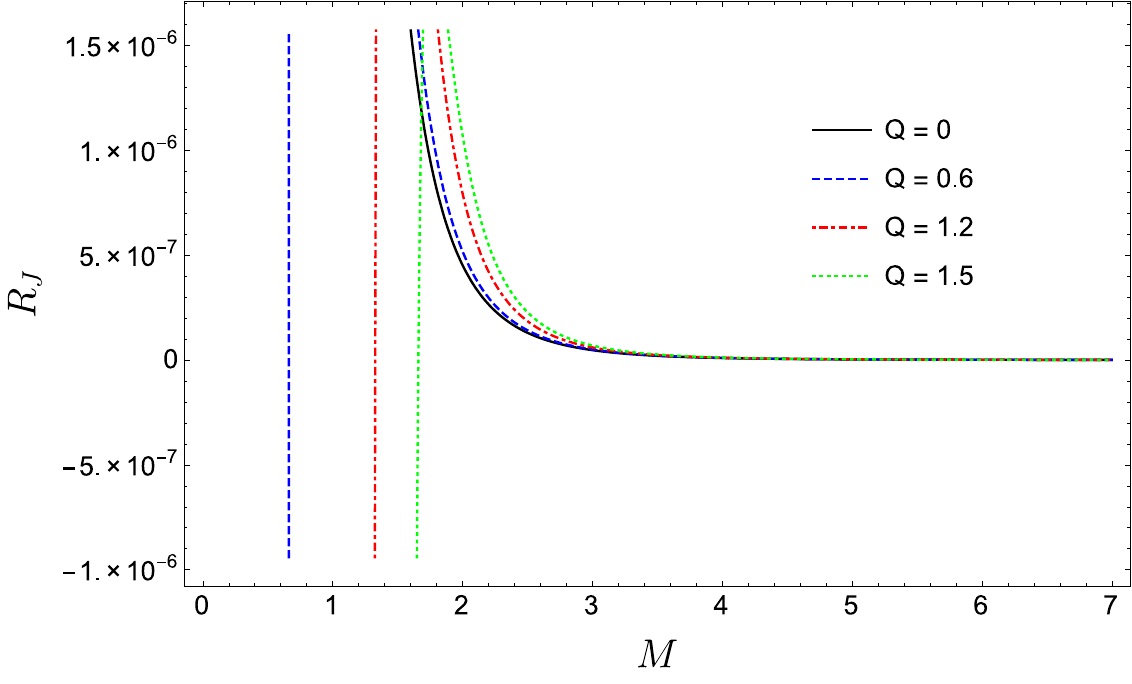}
             \vspace{3mm}
                  \includegraphics[width=8.3cm]{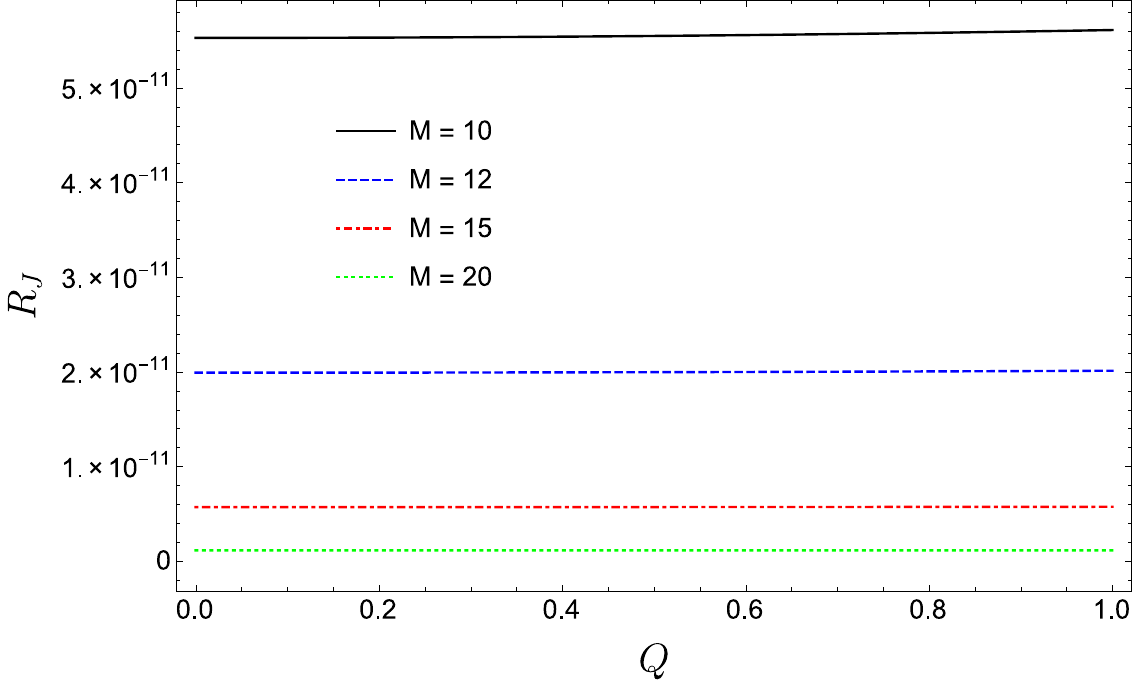}
        \caption{Ruppeneir curvature of Tsallis KNBH for fixed $J$ and fluctuating $(M,Q)$. We set $\delta=0.8$. }
        \label{Ruppe1bis}
    \end{center}
\end{figure}

On the other hand, as $\delta$ approaches unity ($\delta=0.8$ in Fig.~\ref{Ruppe1bis}), $R_J$ is mostly positive, but there is a regime of negative values for sufficiently small $M$ (see both the upper and middle panels). Furthermore, $R_J$ asymptotically vanishes, which means that interactions are balanced in this phase. The standard result of~\cite{Rup2} is recovered in the Boltzmann limit $\delta=1$, where, in fact, $R_J$ is mainly positive and takes negative values near $Q/M=1$.
Regardless of $\delta$, one can verify that $R_J=0$ along the line $J=0$, consistently with the result of~\cite{Amen}. As before, by varying $Q$ while keeping $M$ fixed,  $R_J$ remains nearly constant (lower panel).

\subsubsection{$(M, J)$ fluctuating, $Q$ fixed}

\begin{figure}[t]
    \begin{center}
        \includegraphics[width=9.3cm]{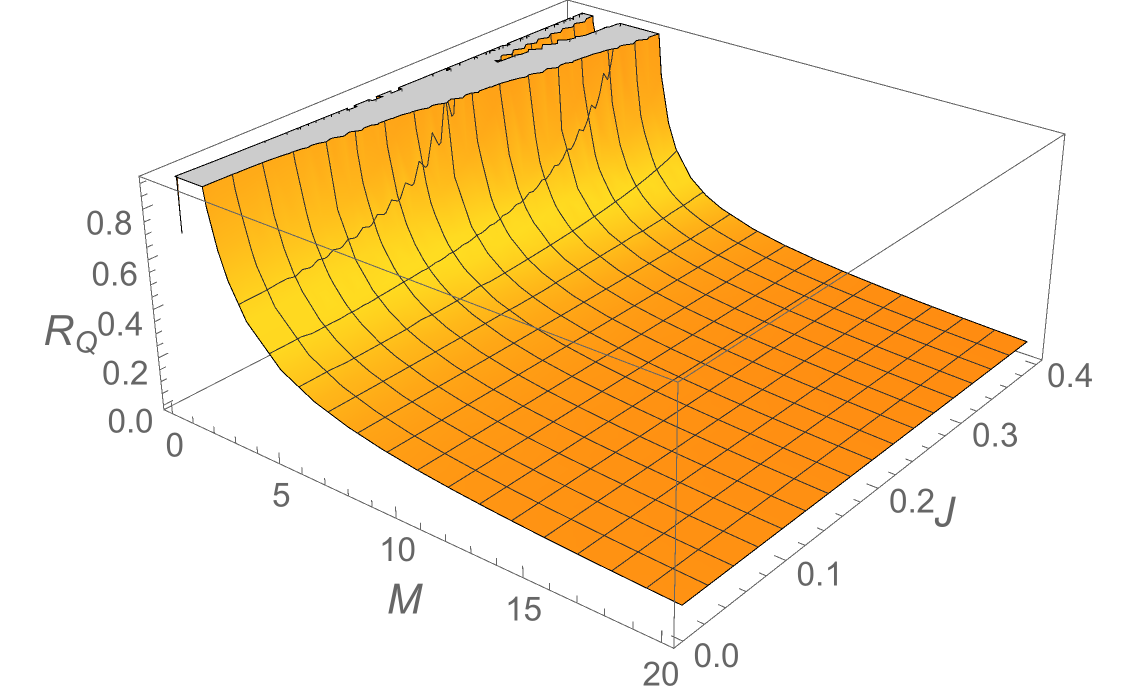}
        
        \vspace{4mm}
          \includegraphics[width=8.3cm]{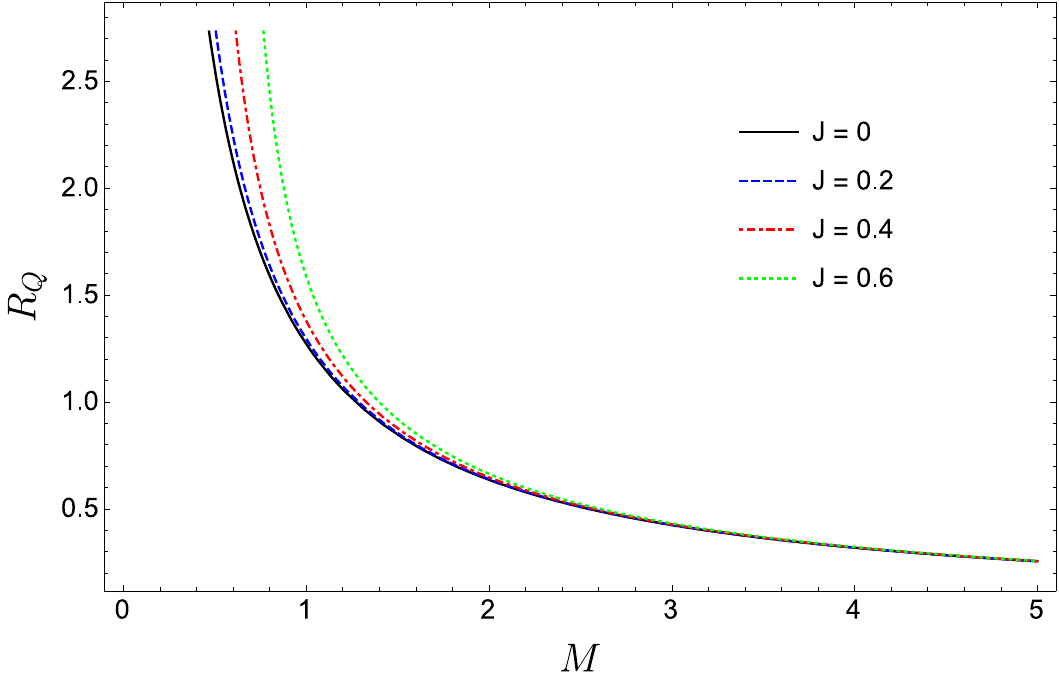}
          \vspace{3mm}
                   \includegraphics[width=8.3cm]{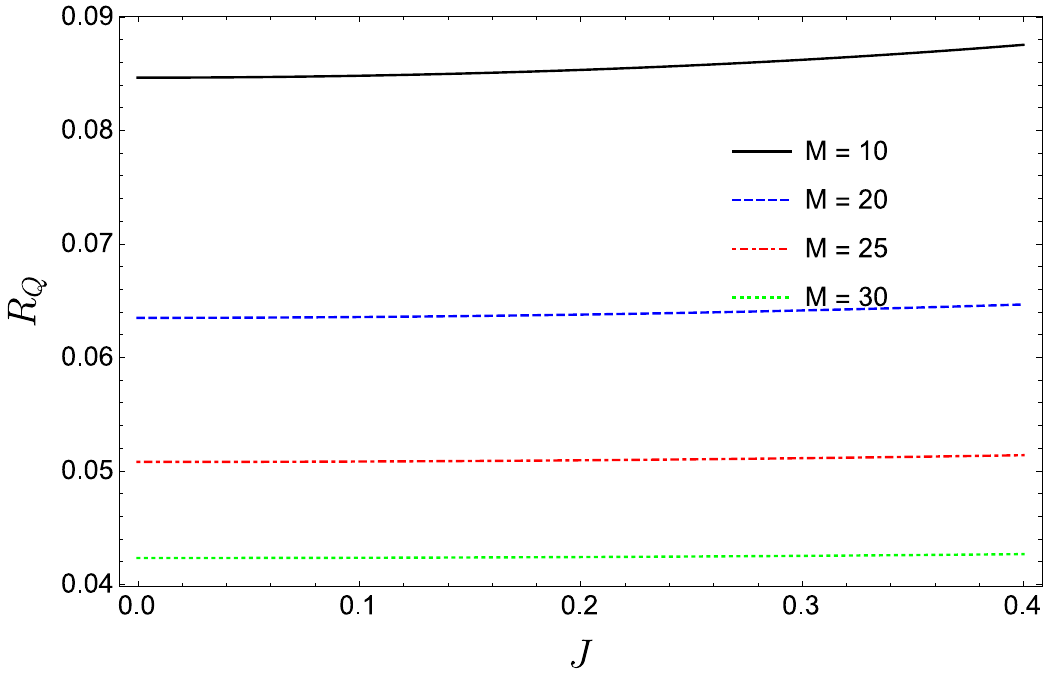}
        \caption{Ruppeneir curvature of Tsallis KNBH for fixed $Q$ and fluctuating $(M,J)$. We set $\delta=0.5$.}
        \label{Ruppe2}
    \end{center}
\end{figure}

\begin{figure}[t]
    \begin{center}
        \includegraphics[width=8.5cm]{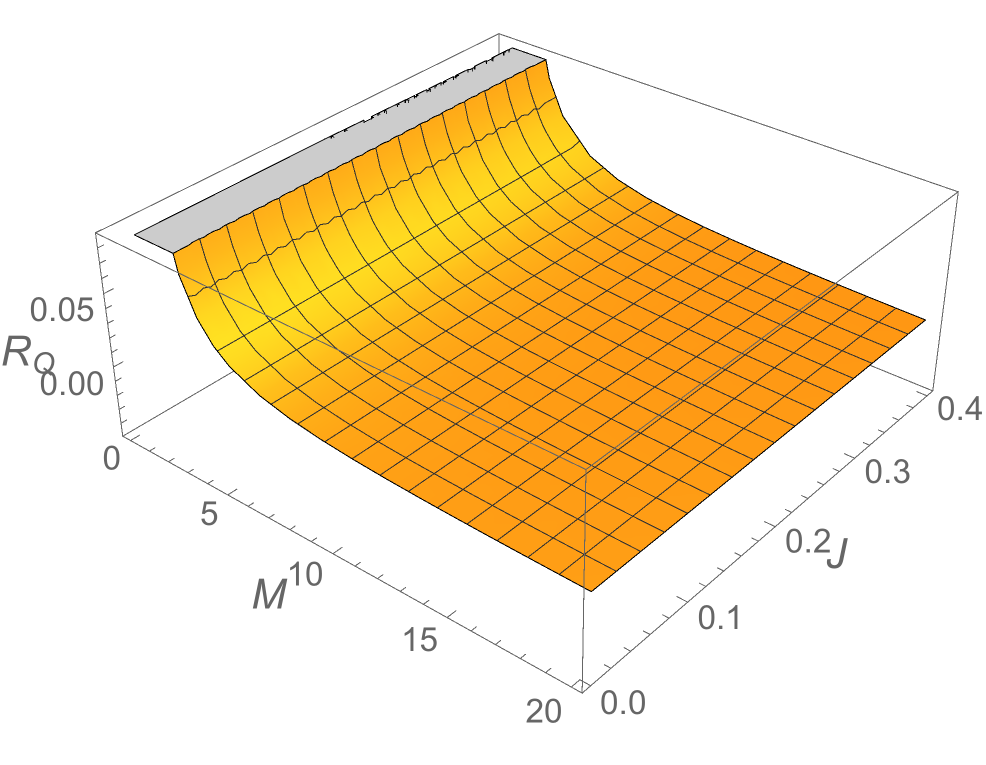}
        
        \vspace{4mm}
          \includegraphics[width=8.3cm]{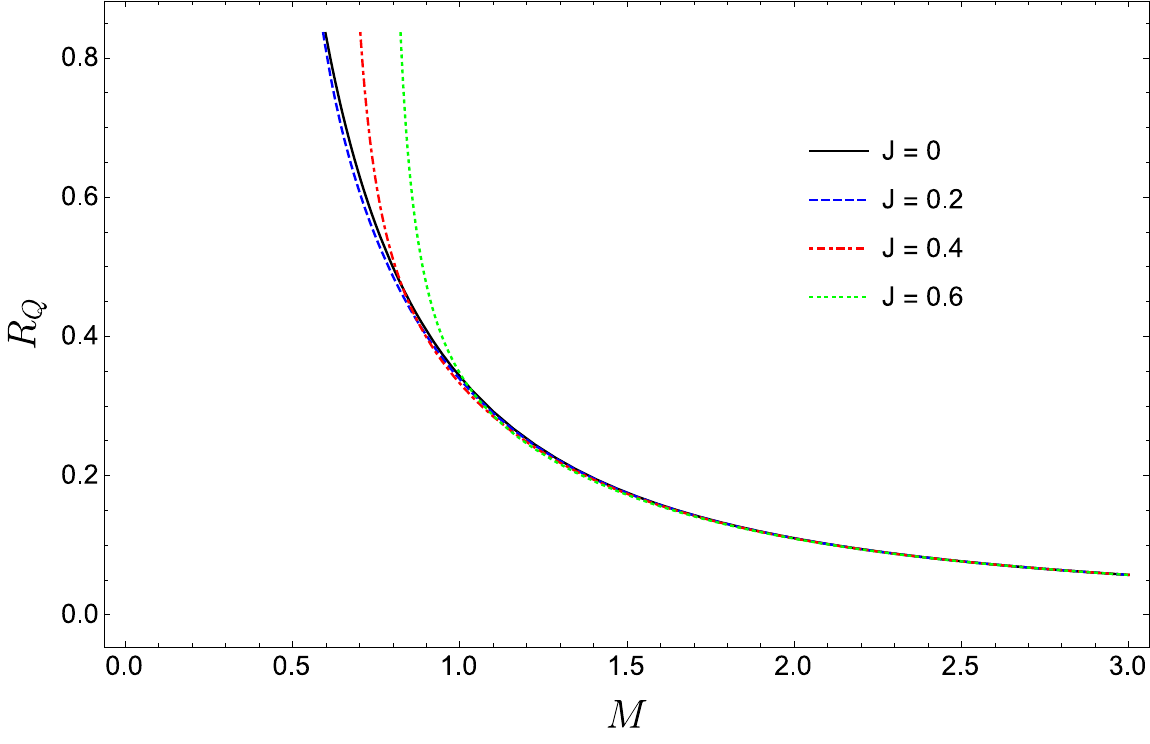}
          \vspace{3mm}
                   \includegraphics[width=8.3cm]{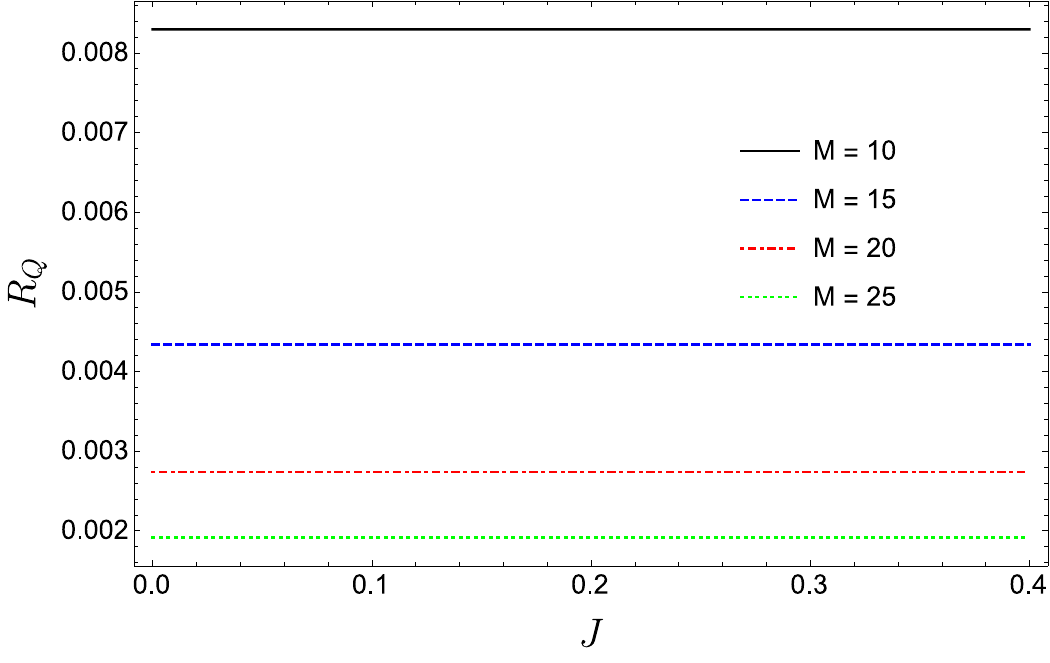}
        \caption{Ruppeneir curvature of Tsallis KNBH for fixed $Q$ and fluctuating $(M,J)$. We set $\delta=0.8$.}
        \label{Ruppe2bis}
    \end{center}
\end{figure}

In this case, we assume that $(M, J)$ fluctuate at fixed $Q$, which gives
\be
ds_R^2=g_{MM}\left(\Delta M\right)^2+2g_{MJ} \left(\Delta M\right) \left(\Delta J\right)+g_{JJ}\left(\Delta J\right)^2\,.
\ee

The curvature $R_Q$ is plotted in Fig.~\ref{Ruppe2} for $\delta=0.5$ and  Fig.~\ref{Ruppe2bis} for $\delta=0.8$. In both cases, $R_Q$ behaves similar to the standard Botlzmann framework~\cite{Rup2}, namely it is real and positive everywhere in the physical regime, and vanishes for $M$ large enough (see the middle panels in Fig.~\ref{Ruppe2} and Fig.~\ref{Ruppe2bis}). On the other hand, for fixed $M$ and varying $J$, $R_Q$ is nearly constant for the selected range of values of the angular momentum (see the lower panels).

\subsection{R\'enyi entropy}
For the case of R\'enyi entropy, we use the definition~\eqref{RenEnt}.
However, due to cumbersome technicalities, we are only able to carry out approximate computations to the leading order in
R\'enyi parameter.

\subsubsection{$(M, Q)$ fluctuating, $J$ fixed}
\begin{figure}[t]
    \begin{center}
        \hspace{-2mm}\includegraphics[width=9.2cm]{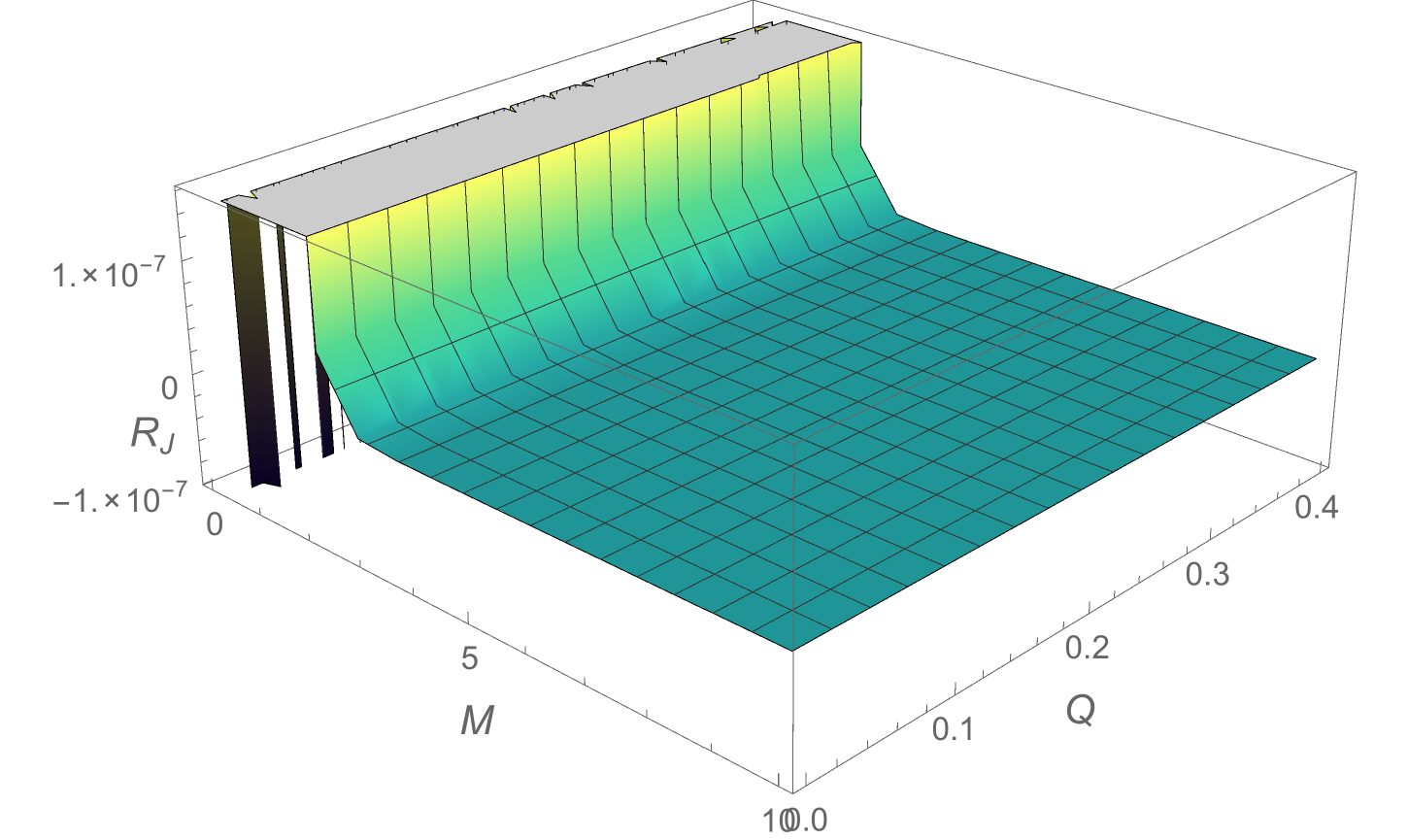}
        
        \vspace{4mm}
        \includegraphics[width=8.3cm]{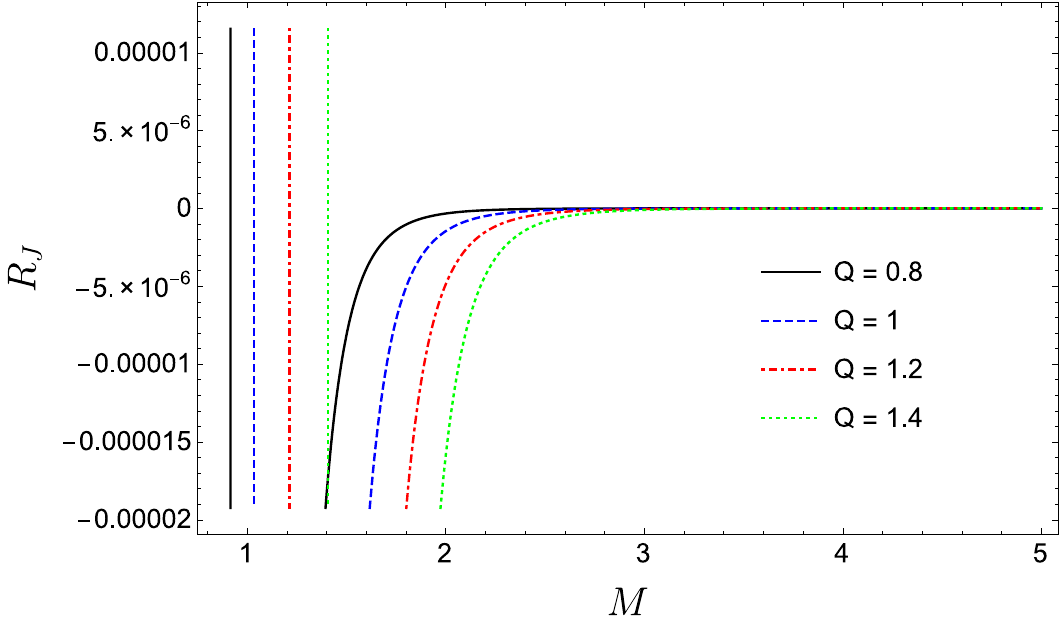}
        \vspace{3mm}
          \includegraphics[width=8.3cm]{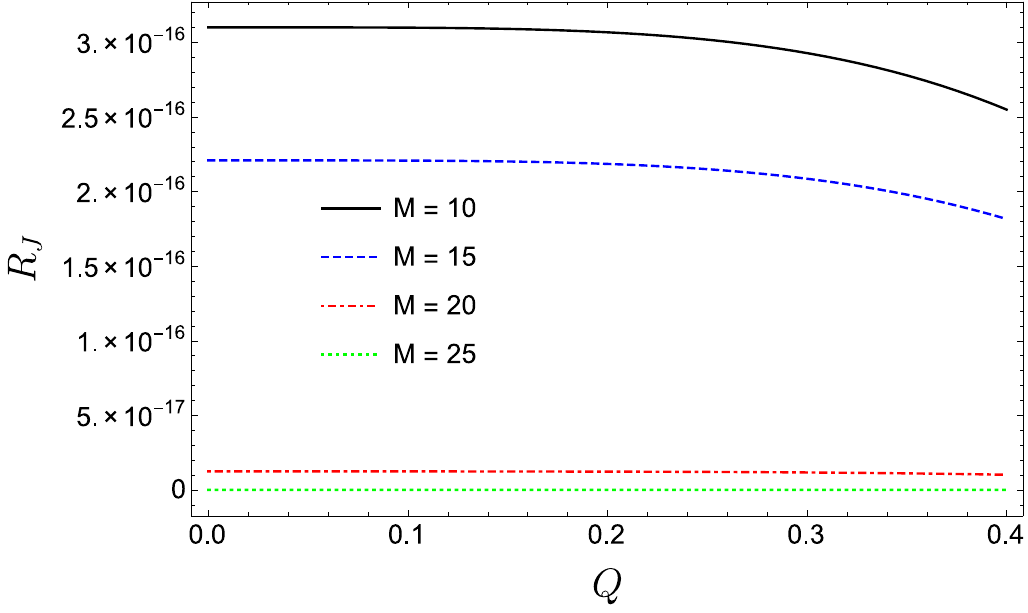}
        \caption{Ruppeneir curvature of  R\'enyi KNBH for fixed $J$ and fluctuating $(M,Q)$. We set $\lambda\sim\mathcal{O}(10^{-4})\ll\lambda_c$.}
        \label{Ruppe3}
    \end{center}
\end{figure}

By plugging Eq.~\eqref{RenEnt} into the line element~\eqref{met1},
we obtain the scalar curvature displayed in Fig.~\ref{Ruppe3}. It can be seen from both the upper and middle panels that $R_J$ drops to negative values in the very small $M$ regime, where microinteractions manifest a prevailing attractive nature, and vanishes asymptotically, where interactions are balanced.
As expected, because of the very small value of $\lambda$ we are considering, such a  behavior resembles the standard one in the Boltzmann regime~\cite{Rup2}. On the other hand, the curvature remains  nearly constant for fixed $M$ and varying $Q$ for the selected range of black hole charge (lower panel).

\subsubsection{$(M, J)$ fluctuating, $Q$ fixed}

\begin{figure}[t]
    \begin{center}
        \hspace{-2mm}\includegraphics[width=9.2cm]{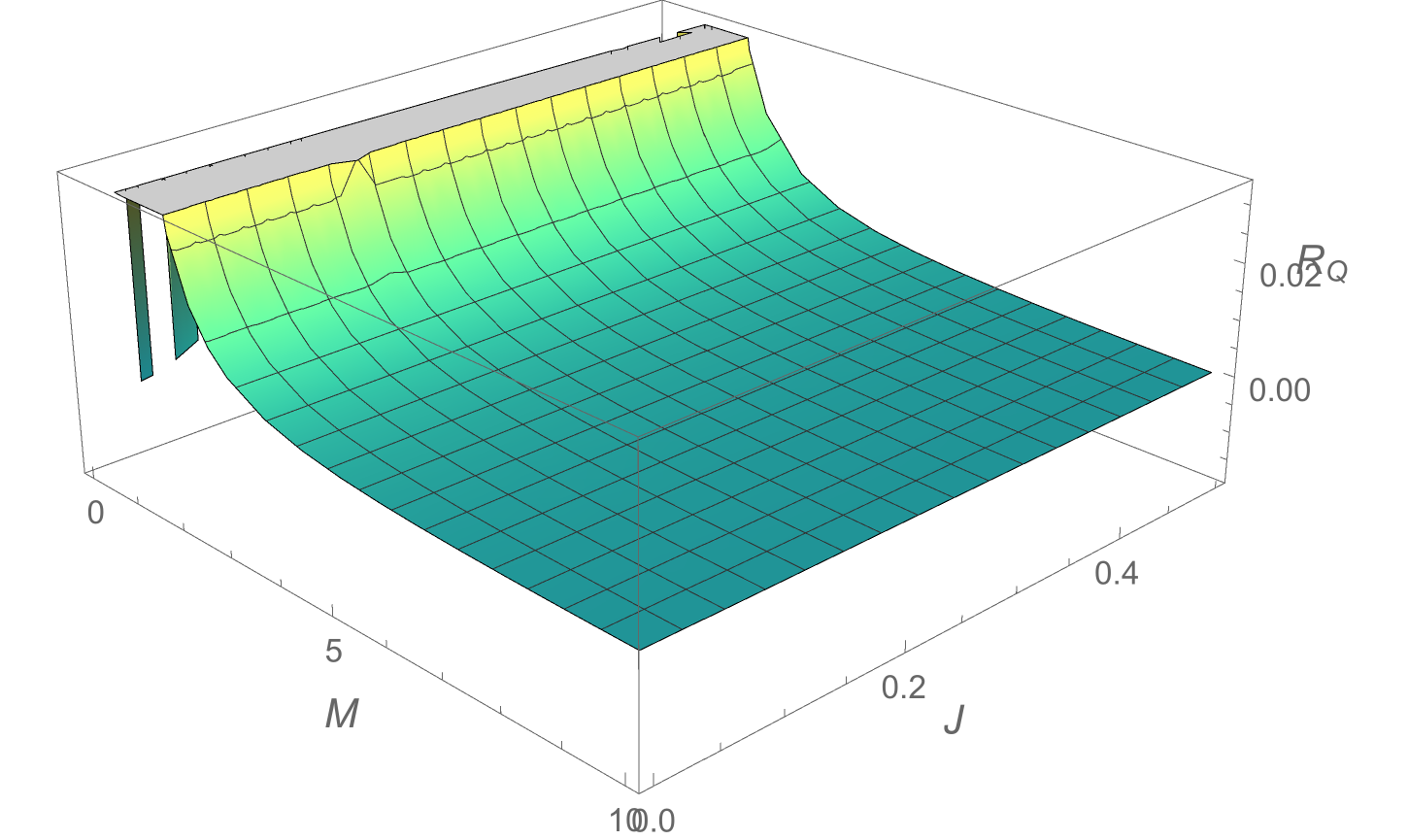}
        
           \vspace{4mm}
        \includegraphics[width=8.1cm]{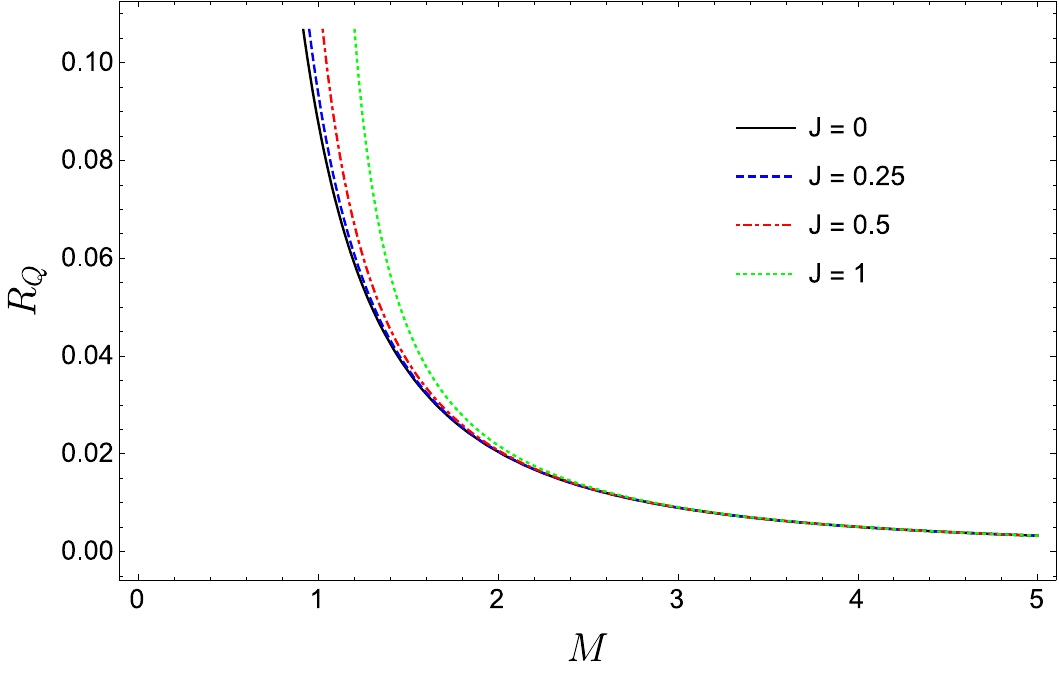}
        \vspace{3mm}
          \includegraphics[width=8.3cm]{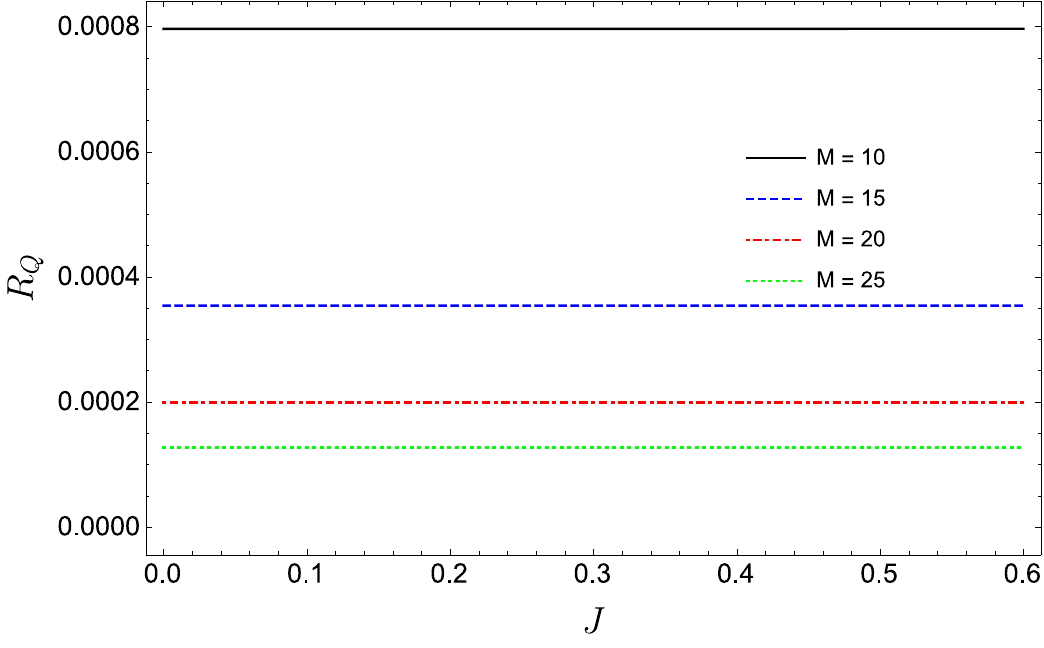}
        \caption{Ruppeneir curvature of  R\'enyi KNBH for fixed $Q$ and fluctuating $(M,J)$. We set $\lambda\sim\mathcal{O}(10^{-4})\ll\lambda_c$.}
        \label{Ruppe4}
    \end{center}
\end{figure}
For $(M, J)$ fluctuating and $Q$ fixed, the Ruppeneir scalar
curvature is plotted in Fig.~\ref{Ruppe4}, which shows that $R_Q$
takes positive values everywhere in the physical regime and
approaches zero for large enough masses (see both upper and middle panels). Once again, this behavior is
similar to the Boltzmann thermodynamics~\cite{Rup2}.
A nearly constant behavior is instead exhibited for fixed $M$ and varying angular momentum $J$ (lower panel).

\section{CONCLUSIONS}
In this paper, we have explored thermodynamics of Kerr-Newman
black holes described by nonextensive Tsallis and  R\'enyi entropy
functions. The issue of thermodynamic and stability properties in
this framework is much complicated than the standard Boltzmann
case, due to the dependence of the entropy expressions on
additional parameters. We use microcanonical and canonical
ensembles to comprehensively analyze thermodynamic stability of
the obtained models. Throughout our study, we have also considered
the behavior of the standard KNBH to compare with our presented
model.

The novel results we addressed in this work can be summarized as
follows. In the Tsallis approach, we have found that thermodynamic
and stability properties mainly depend on the $ \delta $
parameter. In particular, for $ \delta>0.5 $ the behavior of the
temperature versus mass is completely similar to the standard
model and has a global maximum. On the other hand, for $\delta<
0.5$, the temperature is a monotonically increasing function that
blows up asymptotically, while for $\delta=0.5$ it approaches a
constant value for $M$ large enough. Therefore, large deviations
of $\delta$ from unity would have a non-trivial impact on thermal
properties of black holes, with higher values of mass
corresponding to higher temperatures. The ensuing phenomenology is
far from the the conventional Botlzmann picture, where the more
massive a black hole, the colder it is.

As part of the thermodynamic stability analysis, we have
considered both the microcanonical and canonical ensembles, and
plotted the stability curves of the KNBH based on both the
standard Boltzmann and Tsallis entropies, for comparison. We have
observed that there is no change in stability for isolated black
holes in either. In other words, isolated KNBH with Tsallis
entropy behave similar to the standard Boltzmann case $
(\delta=\gamma=1) $ and are thermodynamically stable against
axially symmetric perturbations.

In the canonical ensemble, a bath of thermal radiation surrounds
the black holes. In this context, our findings show that the value
of the $ \delta $ parameter plays a crucial role in determining
the stability.
When $ \delta>0.5 $, there is a turning point in $-M(\beta) $
curves, indicating stability change at this point: in particular,
smaller black holes with positive slope (before the turning point)
are more stable than large black holes with negative slope (after
the turning point). While in the  $ \delta\leq 0.5 $ case, we have
observed that there is no turning point in $ -M(\beta) $ curve,
and as a result, there is no stability change.

Interestingly enough, the above results find confirmation by
application of the traditional Hessian, which show that the
turning point in the $ -M(\beta) $ curve for $ \delta>0.5 $
corresponds to the same point of divergence in the heat capacity
diagram. Aa a result, this discontinuity causes two categories:
stable smaller black holes with $ C_{J,Q} >0$ and unstable larger
black holes $ C_{J,Q}<0 $. For  $ \delta\leq 0.5 $, no change of
stability occur and the black hole is stable for any mass.

In parallel, we have also considered the nonextensive R\'enyi entropy to study thermodynamic properties of the KNBH.
As expected, in the limit of $ \lambda=0 $, the standard Boltzmann thermodynamics is restored.
By analyzing the temperature curves, we have found that there
are a local maximum and a local minimum as far as $\lambda$ lies below
a critical value $\lambda_c$ depending on the model parameters. These stationary points converge into an inflection point as $\lambda$ increases to $\lambda_c$, while $T$ increases monotonically for $\lambda>\lambda_c$. It has been demonstrated that isolated KNBH associated with R\'enyi entropy are stable, like in the Tsallis approach.

Our studies in the canonical ensemble reveal that for $0<\lambda<\lambda_c$, the $ -M(\beta) $ curve has two turning points where the stability changes.
These points correspond to the first-order small black hole/large black hole phase transition (divergence points)
in the heat capacity curve. Three regimes can be distinguished in this case: smaller and large stable black holes, where heat capacity is positive, intermediate unstable black holes with a negative heat capacity.
However, with the increase of the $ \lambda$ value, these instability points disappear and the heat capacity takes positive values for all masses. Such peculiar behaviors are confirmed by the study of the free energy against the temperature. Indeed, the free energy exhibits the characteristic swallow-tail shape as far as $\lambda$ lies below the critical value, which is typical of the first order small/large black hole transition. By increasing $\lambda$ at the critical value,  the swallow-tail behavior disappears, which suggests that the phase transition becomes second order. Above $\lambda_c$, there is no stability change.

We have finally explored black hole substructure by use of geometrothermodynamic Ruppeneir formalism, which allows to
deduce information on the nature of microinteractions by looking at the sign of the scalar curvature of Ruppeneir metric. Calculations have been performed by considering either $(M.Q)$ (case 1) or $(M,J)$ (case 2) as fluctuating variables. In Tsallis framework and for the case 1, we have shown that the curvature is positive and increases for increasing mass as far as $\delta\le0.5$, which is indicative of prevailing repulsive interactions. On the other hand, for $\delta>0.5$  the curvature is mostly positive with a regime of negative values (attractive interactions) for small enough mass. This behavior resembles the traditional finding in Boltzmann statistics.
Similarities with Boltzmann entropy appear more evident in case 2, where the curvature is real and positive everywhere in the physical regime and vanishes at sufficiently large mass for any $\delta$.

Concerning R\'enyi entropy, computations have been performed
to the leading order in $\lambda$ due to technicalities. In spite of this assumption, Ruppeneir curvature still provides an interesting perspective on the internal interactions. For the case 2, it is always positive and asymptotically vanishes. The same qualitative behavior occurs in the case 1, except for small $M$, where the curvature drops to negative values.

Further aspects are yet to be investigated: first, it would be suggestive to
extend the present study, so as to include global monopoles.
In fact, phase transitions in the early Universe can
originate topological defects like global monopoles~\cite{Kib}
Gravitational fields related with these solutions are responsible for fluctuations in the microwave background radiation,
which are the seeds of galaxy clusters and are potentially observable.
Black holes with global monopoles have been largely considered
in literature (see~\cite{Mon1,Mon2,Mon3} and references therein), due to their richer phenomenology and topological structure comparing to Schwarzschild black holes. Hence, it is interesting to explore the interplay of monopole effects and nonextensive entropies.
Furthermore, one could develop thermodynamics of KNBH in other generalized entropy frameworks. In this line, Kaniadakis statistics~\cite{Kania} has attracted a great deal of attention as an effort to incorporate relativistic symmetries in the classical Boltzmann theory (see~\cite{LucReview} for a recent review of applications in cosmological and gravitational contexts).
Work along these directions is under active consideration and will
be presented elsewhere.



\begin{thebibliography}{99}

\bibitem{Bekenstein}
J.D. Bekenstein, Phys. Rev. D {\bf 7}, 2333 (1973).

\bibitem{Bek2}
J.D. Bekenstein, Phys. Rev. D {\bf 9} 3292  (1974).

\bibitem{Hawking;1974} S.W. Hawking, Nature 248, \textbf{30} (1974).

\bibitem{Hawking1} S.W. Hawking, Commun. Math. Phys. {\bf 43}, 199 (1975).

\bibitem{Hawking2} S.W. Hawking, Phys. Rev. D {\bf 13}, 191 (1976).

\bibitem{Bardeen}
J.~M.~Bardeen, B.~Carter and S.~W.~Hawking,
Commun. Math. Phys. \textbf{31}, 161 (1973).

\bibitem{Gibbs} J.W. Gibbs, \textit{Elementary Principles of Statistical Mechanics} (Dover, New York, 1960).

\bibitem{Landsberg1} P.T. Landsberg, J. Stat. Phys. \textbf{35}, 159 (1984).


\bibitem{Landsberg2} P.T. Landsberg, D. Tranah, Phys. Lett. A {\bf78}, 219 (1980).

\bibitem{Tsallis} C. Tsallis, L.J.L. Cirto, Eur. Phys. J. C {\bf73}, 2487 (2013).

\bibitem{Pavon} D. Pav\'on, J.M. Rubí, Gen. Rel. Grav. {\bf18}, 1245 (1986).

\bibitem{Maddox} J. Maddox, Nature {\bf365}, 103 (1993).

\bibitem{Pesci} A. Pesci, Class. Quant. Grav. {\bf24}, 2283 (2007).

\bibitem{Davies} P.C.W. Davies, Proc. R. Soc. Lond. A {\bf353}, 499 (1977).

\bibitem{Kaburaki1} O. Kaburaki, I. Okamoto, J. Katz, Phys. Rev. D {\bf 47}, 2234 (1993).

\bibitem{Kaburaki1bis}
 O. Kaburaki, Gen. Relativ. Gravit. {\bf28}, 843 (1996).

\bibitem{Kaburaki2}O. Kaburaki, I. Okamoto, J. Katz, Class. Quant. Grav. {\bf10}, 1323 (1993).


\bibitem{Czinner} V.G. Czinner, Int. J. Mod. Phys. D {\bf 24},  1542015 (2015).

\bibitem{Czinner2}
V. G. Czinner, H. Iguchi, Phys. Lett. B, {\bf752}, 306 (2016).

\bibitem{Czinner3}
V.G. Czinner, H. Iguchi, Universe {\bf3}, 14 (2017).

\bibitem{Czinner4}
V.G. Czinner, H. Iguchi, Eur. Phys. J. C  {\bf 77}, 892 (2017).

\bibitem{Azreg}
M.~Azreg-A\"\i{}nou and M.~E.~Rodrigues,
JHEP \textbf{09}, 146 (2013).



\bibitem{Biro} T. S. Bir\'o and V. G. Czinner, Phys. Lett. B {\bf726}, 861 (2013).

\bibitem{Nojiri-non1}S. Nojiri, S.D. Odintsov, V. Faraoni, Phys. Rev. D {\bf104}, 084030 (2021).

\bibitem{Nojiri-non2}
S. Nojiri, S.D. Odintsov, V. Faraoni, Int. J. Geom. Methods
Mod. Phys. \textbf{19}, 2250210 (2022).






\bibitem{Luciano1} G.G. Luciano, M. Blasone, Phys. Rev. D \textbf{104}, 045004 (2021).

\bibitem{Vagnozzi}
S.~Vagnozzi, R.~Roy, Y.~D.~Tsai, L.~Visinelli, M.~Afrin, A.~Allahyari, P.~Bambhaniya, D.~Dey, S.~G.~Ghosh and P.~S.~Joshi, \textit{et al.}
Class. Quant. Grav. \textbf{40}, 165007 (2023).

\bibitem{Morad2} H. Moradpour, Int. J. Theor. Phys. {\bf55}, 4176 (2016).

\bibitem{Nunes}
R.~C.~Nunes, E.~M.~Barboza, Jr., E.~M.~C.~Abreu and J.~A.~Neto,
JCAP \textbf{08}, 051 (2016).

\bibitem{Komatsu} N. Komatsu, Eur. Phys. J. C {\bf77}, 229 (2017).

\bibitem{Nojiri-non3} S. Nojiri, S.D. Odintsov, V. Faraoni, Phys. Rev. D {\bf105}, 044042 (2022).

\bibitem{Ghaff1}
S.~Ghaffari, H.~Moradpour, A.~H.~Ziaie, F.~Asghariyan, F.~Feleppa and M.~Tavayef,
Gen. Rel. Grav. \textbf{51}, 93 (2019).


\bibitem{Tavayef:2018xwx}
M.~Tavayef, A.~Sheykhi, K.~Bamba and H.~Moradpour,
Phys. Lett. B \textbf{781}, 195 (2018).

\bibitem{Saridakis:2018unr}
E.~N.~Saridakis, K.~Bamba, R.~Myrzakulov and F.~K.~Anagnostopoulos,
JCAP \textbf{12}, 012 (2018).

\bibitem{Nojiri:2019skr}
S.~Nojiri, S.~D.~Odintsov and E.~N.~Saridakis,
Eur. Phys. J. C \textbf{79}, 242 (2019).

%
\bibitem{Jizba}
P.~Jizba, G.~Lambiase, G.~G.~Luciano and L.~Petruzziello,
Phys. Rev. D \textbf{105}, L121501 (2022).

\bibitem{LucianoGen}
G.~G.~Luciano and J.~Gin\'e,
Phys. Dark Univ. \textbf{41}, 101256 (2023)

\bibitem{Boulka}
N.~Boulkaboul,
Phys. Dark Univ. \textbf{40}, 101205 (2023).

\bibitem{Morad3}
S.~Jalalzadeh, H.~Moradpour and P.~V.~Moniz,
Phys. Dark Univ. \textbf{42}, 101320 (2023).

\bibitem{Capoz1}
M. Zubair and L. R. Durrani, Chin. J. Phys. \textbf{69}, 153
(2021).

\bibitem{Capoz2}
Y.~Aditya, S.~Mandal, P.~K.~Sahoo and D.~R.~K.~Reddy,
Eur. Phys. J. C \textbf{79}, 1020 (2019).

\bibitem{Capoz3}
S.~Capozziello and M.~De Laurentis,
Phys. Rept. \textbf{509}, 167 (2011).

\bibitem{Capoz4}
S.~Ghaffari, G.~G.~Luciano and S.~Capozziello,
Eur. Phys. J. Plus \textbf{138}, 82 (2023).

\bibitem{Capoz5}
S.~Capozziello and M.~Francaviglia,
Gen. Rel. Grav. \textbf{40}, 357 (2008).

\bibitem{Capoz6}
A.~Sarkar and S.~Chattopadhyay,
Int. J. Geom. Meth. Mod. Phys. \textbf{18}, 2150148 (2021).

\bibitem{LucianoBar}
G.~G.~Luciano,
Phys. Dark Univ. \textbf{41}, 101237 (2023).

\bibitem{Luciano}
G.~G.~Luciano and A.~Sheykhi,
Phys. Dark Univ. \textbf{42}, 101319 (2023)

\bibitem{Luciano:2023bai}
G.~G.~Luciano and E.~Saridakis,
[arXiv:2308.12669 [gr-qc]].

\bibitem{Jawad}
A. Jawad and S. R. Fatima, Nucl. Phys. B \textbf{976}, 115697
(2022).

\bibitem{Yassine}
Y.~Sekhmani, J.~Rayimbaev, R.~Myrzakulov and D.~J.~Gogoi,
[arXiv:2311.02448 [gr-qc]].

\bibitem{Cimidiker:2023kle}
I.~Cimidiker, M.~P.~Dabrowski and H.~Gohar,
Class. Quant. Grav. \textbf{40}, 145001 (2023).

\bibitem{Morad1} S. Rani, A. Jawad, H. Moradpour, A. Tanveer, Eur. Phys. J. C \textbf{82},  713 (2022).

\bibitem{Villalba}
F.~D.~Villalba, A.~F.~Vargas, E.~Contreras and P.~Bargue\~no,
Gen. Rel. Grav. \textbf{52}, 87 (2020).

\bibitem{Rup}
G. Ruppeiner, Phys. Rev. A \textbf{20}, 1608 (1979).

\bibitem{Poincare} H. Poincar\'e, Acta Math. \textbf{7}, 259 (1885).

\bibitem{Ruiz:2019aoh}
O.~Ruiz, U.~Molina and P.~Viloria,
J. Phys. Conf. Ser. \textbf{1219}, 012016 (2019).

\bibitem{Wei}
F. Weinhold, J. Chem. Phys. \textbf{63}, 2479 (1975).


\bibitem{Cai}
R. G. Cai and J. H. Cho, Phys. Rev. D \textbf{60}, 067502 (1999).

\bibitem{PRL}
S. W. Wei and Y. X. Liu, Phys. Rev. Lett. \textbf{115}, 111302 (2015).

\bibitem{WeiLiu}
S. W. Wei, Y. X. Liu and R. B. Mann, Phys. Rev. Lett. \textbf{123}, 071103 (2019).

\bibitem{Xu}
Z. M. Xu, B. Wu and W. L. Yang, Phys. Rev. D \textbf{101}, 024018 (2020).

\bibitem{Ghosh}
A. Ghosh and C. Bhamidipati, Phys. Rev. D \textbf{101}, 106007
(2020).

\bibitem{Rup2}
G.~Ruppeiner,
Phys. Rev. D \textbf{78}, 024016 (2008).

\bibitem{Amen}
J.~E.~Aman, I.~Bengtsson and N.~Pidokrajt,
Gen. Rel. Grav. \textbf{35}, 1733 (2003).

\bibitem{Kib}
T. W. B. Kibble, J. Phys. A \textbf{9} 1387 (1976).

\bibitem{Mon1}
J. L. Jing, H.W. Yu, Y.J. Wang, Phys. Lett. A \textbf{178}  59 (1993).

\bibitem{Mon2}
 H.W. Yu, Nucl. Phys. B \textbf{430}  427 (1994).

 \bibitem{Mon3}
 X.~z.~Li and J.~g.~Hao,
Phys. Rev. D \textbf{66}, 107701 (2002).

\bibitem{Kania}
G. Kaniadakis, Physica A \textbf{296}, 405 (2001).

\bibitem{LucReview}
G.~G.~Luciano,
Entropy \textbf{24}, 1712 (2022).

\end{thebibliography}
\end{document}